\documentstyle[12pt]{article}

\renewcommand{\theequation}{\thesection.\arabic{equation}}

\newcommand \beq{\begin{eqnarray}}
\newcommand \eeq{\end{eqnarray}}
\font\tennbf=cmbx12 \newfam\nbffam
\textfont\nbffam=\tennbf
\def\nbf{\fam\nbffam\tennbf}
\font\tennrm=cmr12 \newfam\nrmfam
\textfont\nrmfam=\tennrm
\def\nrm{\fam\nrmfam\tennrm}

\def\bfgrad{\mbox{\boldmath$\grad$}}

\def\grad{\nabla}
\def\del{\partial}

\def\frac#1#2{{#1 \over #2}}

\def\half{\ifinner {\scriptstyle {1 \over 2}}
   \else {1 \over 2} \fi}

\def\simge{\mathrel{%
   \rlap{\raise 0.511ex \hbox{$>$}}{\lower 0.511ex \hbox{$\sim$}}}}
\def\simle{\mathrel{
   \rlap{\raise 0.511ex \hbox{$<$}}{\lower 0.511ex \hbox{$\sim$}}}}

\def\slashchar#1{\setbox0=\hbox{$#1$}
   \dimen0=\wd0
   \setbox1=\hbox{/} \dimen1=\wd1
   \ifdim\dimen0>\dimen1
      \rlap{\hbox to \dimen0{\hfil/\hfil}}
      #1
   \else
      \rlap{\hbox to \dimen1{\hfil$#1$\hfil}}
      /
   \fi}

\def\subrightarrow#1{
  \setbox0=\hbox{
    $\displaystyle\mathop{}
    \limits_{#1}$}
  \dimen0=\wd0
  \advance \dimen0 by .5em
  \mathrel{
    \mathop{\hbox to \dimen0{\rightarrowfill}}
       \limits_{#1}}}                           

\def\journal#1#2#3#4{\ {#1}{\bf #2} ({#3})\  {#4}}

\def\NPB{\journal{Nucl.\ Phys.\ {\bf B}}}

\def\PLB{\journal{Phys.\ Lett.\ {\bf B}}}

\def\PRD{\journal{Phys.\ Rev.\ {\bf D}}}
\def\PRL{\journal{Phys.\ Rev.\ Lett.}}

\def\RevModPhys{\journal{Rev.\ Mod.\ Phys.}}

\begin{document}

\begin{titlepage}
\begin{flushright} {Saclay-T95/034} \\ {IPNO/TH 95-17} \\ {IP/BBSR/95-13}
\end{flushright}
\vspace*{0.5cm}
\begin{center}
\baselineskip=13pt {\Large ON THE SCREENING OF STATIC ELECTROMAGNETIC}
\vskip0.3cm {\Large FIELDS IN HOT QED PLASMAS} \vskip0.5cm Jean-Paul
BLAIZOT\footnote{CNRS} \\ {\it Service
de Physique Th\'eorique\footnote{Laboratoire de la Direction des Sciences
de
la Mati\`ere du Commissariat \`a l'Energie Atomique}, CE-Saclay \\ 91191
Gif-sur-Yvette, France}
\vskip0.3cm
 Edmond IANCU \\ {\it Division de Physique
Th\'eorique\footnote{Unit\'e de Recherche des Universit\'es Paris XI et Paris
VI associ\'ee au CNRS}, I.P.N.-Orsay\\
         91406 Orsay, France}
\vskip0.1cm
 and \\ Rajesh R. PARWANI \\ {\it Institute of Physics\\
Bhubaneswar 751005, India}\\
\vskip0.5cm

\end{center}

\vskip 1.5cm
\begin{abstract}
 We study the screening of static magnetic and electric
fields in massless quantum electrodynamics (QED) and  massless scalar
electrodynamics (SQED) at temperature $T$. Various exact relations for
the static polarisation tensor are first reviewed and then verified
perturbatively to fifth order (in the coupling)
in QED and fourth order in SQED, using different resummation
techniques. The magnetic and electric screening masses squared,
 as defined through the pole of the static
propagators, are also calculated to fifth order in QED and fourth order in
SQED, and their gauge-independence and renormalisation-group
invariance is checked. Finally, we provide arguments
for the vanishing of the magnetic
mass to all orders in perturbation theory.
\end{abstract}

\vskip 3cm


\end{titlepage}

\setcounter{equation}{0}
\section{Introduction}

The simplest and probably best known manifestation of many-body effects in
electromagnetic plasmas is Debye screening : the usual
Coulomb potential between two static charges in a vacuum transforms, in the
presence of a plasma, into a Yukawa potential, $V(r) \sim {{\nrm e}^
{-m_D r}/r}$.
 The scale $m_D$ is called the electric (Debye) screening mass. For a
plasma at high temperature $T$ (much larger than the electron mass), $m_D^2
= {e^2 T^2 / 3}$ to lowest order in the coupling constant\cite{Weldon}.

The above-mentioned relationship between the static potential
and the screening mass is usually established
within the approximation of linear response (see \cite{Kapusta89} and
section 2),  whereby one calculates the
potential between two arbitrarily weak {\it external
static} sources $q_1$ and $q_2$ separated by $\vec{r}$,
\begin{eqnarray}
V(r) = q_1 q_2 \int {d^3 p \over (2 \pi)^3 } {e^{i \vec{p} \cdot \vec{r}}
\over p^2 + \Pi_L(0,p) } \, =
\frac{q_1 q_2}{2\pi r} \int_{-\infty}^\infty {{\nrm d} p \over 2 \pi i }
\, {p\,e^{i{p}{r}} \over p^2 + \Pi_L(0,p) } \, , \label{pot} \end{eqnarray}
with  $\Pi_L(0,p)=-\Pi_{00}(0,p)$,
 $\Pi_{00}(0,p)$ being the static ($p_0=0$) electric polarisation
tensor (our conventions and notations are summarized in section 2).
 The integral over $p$ may be performed
by closing the integration path in the upper half
of the complex $p$ plane. When the external charges are widely
separated ($ r \to \infty$),
the behaviour of the potential is dominated by the
 singularity of $D_{00}(0,p)\equiv - 1/(p^2 + \Pi_L(0,p))$
 which lies closest to the real axis. To leading order, this is a pole
 occuring  for $p^2= \Pi_{00}^{(2)}(0,p\to 0) = - e^2 T^2/3$,
 $ \Pi_{00}^{(2)}$  being the one-loop polarization tensor.

One of the main objectives of this paper is to study, within the
definition (\ref{pot}), corrections to this leading ($ \sim e^2 T^2$)
Debye mass, and in the process clarify several issues
which arise. We shall verify in a perturbative calculation that
 the dominant singularity of $D_{00}$ remains a pole
on the imaginary axis.  This allows us
to define $m_D$ as the solution of \cite{Rebhan93}
\beq\label{debyemass}
m_D^2=-\Pi_{00}(0,p)\left.\right|_{p^2=-m_D^2}.
\eeq
We shall perform the computation in eq.(\ref{debyemass})
up to order $e^5$ in usual (spinor)
electrodynamics  (QED), and up to order $e^4$ in scalar electrodynamics
 (SQED). In both cases, this corresponds
to the first two non-trivial corrections above the  ``hard thermal loop''
approximation\cite{BP90,FT90}, and is an extension of previous lower
order calculations.
As we shall see, the Debye mass obtained in this way
is renormalisation group invariant to the required order.
 General arguments\cite{KKR90}
 also indicate the gauge-independence of
propagator poles and this too will be demonstrated in our
computations. At this point it is worth noting
that,  though we
have introduced the screening mass via the potential above,
the same mass $m_D$ controls the exponential decay of other
interesting gauge-invariant correlators such as that of static
electric fields $\langle E_i(x) E_j(y) \rangle$.

Solving eq.~(\ref{debyemass}) requires the knowledge of
 $ \Pi_{00}(0,p)$ for $p \sim m \sim eT$.
Thus we are led to study first the simpler object
$\Pi_{00}(0,0)$, up to the order of interest.
 In massless QED, this quantity has been already computed
to order $e^5$, but only indirectly, by using the
 relation\cite{Fradkin65,Kapusta89}
\beq\label{derpress}
\Pi_{00}(0,p\to 0) \,=\,-\,e^2\,\frac{\del^2 P}{\del \mu_e^2},\eeq
(where $P$ is  the plasma pressure  and $\mu_e$ is the chemical potential
for the  charged particles), together with an old result for the
pressure\cite{Akhiezer60}.  Notice that,
because of the explicit factor of $e^2$ on the
 right-hand-side (r.h.s.) of eq.~(\ref{derpress}), the
result for the pressure is only required to third order (but for finite
chemical potential). Still, since the derivation of eq.~(\ref{derpress})
is a formal one (see section 2.2),  it is
of interest to verify the identity explicitly in some cases to show that it
is not invalidated, for example, by potential problems such as infrared
divergences. This, we shall do in the case of QED:  we shall compute
the left-hand-side (l.h.s.)
 of (\ref{derpress}) to fifth order (and  at zero $\mu_e$),
and  we shall check that it agrees
with known results for the r.h.s. In the case of
SQED, $\Pi_{00}(0,0)$ has been
previously computed up to order $e^3$\cite{Kalashnikov80}
(see also Ref. \cite{Kraemmer94}); we will present in this
paper the order $e^4$ correction.

Another objective here is to look into the screening of static magnetic fields.
It is known that such  fields are not screened
in ordinary plasmas. Indeed, one may rely on Ward identities, together
with the (exact) Dyson-Schwinger equation to show that
\begin{eqnarray}
\Pi_{ij}(0, p \to 0) = 0 \, , \label{mag} \end{eqnarray}
(see Sect. 2.2 for more details).
However, in order to guarantee the absence
 of perturbative singularities beyond
that at $p^2 =0$ in the correlator of magnetic fields, one needs
the stronger result $ \Pi_{ij}(0, p\to 0) = {\cal O} (p^2)$. This will
be verified explicitly up to fifth (respectively, fourth)
order in perturbation theory for  QED (respectively, SQED).
The perturbative arguments can be  extended
to an all-order proof, to be detailed in Sect. 3.1 in the case of QED.

As in all higher order perturbative calculations at nonzero
temperature, a systematic determination of  $m_D$ requires
a resummation of the large collective plasma effects.
For the calculation of dynamical quantities,
the resummation involves the procedure developed
by Braaten and Pisarski \cite{BP90}
whereby one uses effective vertices and
propagators obtained by dressing the bare quantities with
``hard thermal loops''. The latter are the dominant parts of
one-loop amplitudes and express the effects of
Landau damping,  Debye screening and  collective oscillations
\cite{BP90,FT90,BI}.
However for the computation of
static (zero external energy) Green's functions, like those
to be discussed here, the calculations are most natural and
convenient in the imaginary time formalism, without analytical
continuation to real time. Then the general
 programme of Refs. \cite{BP90} reduces to
the well known Gell-Mann-Bruekner
\cite{GMB} resummation studied many years ago in
the nonrelativistic context and subsequently extended to the
relativistic regime \cite{Akhiezer60,inspect,staticres}.

For static calculations in the
imaginary time formalism (which we use exclusively
in this paper), the resummation concerns only the internal lines with
 zero Matsubara frequencies (also referred as {\it static} lines), and
consists in dressing these lines with the corresponding
screening thermal masses.
 No vertex resummation will be needed: in QED, there are no vertex in which all
lines are soft; in SQED, there are no hard thermal
loops beyond the two-point functions \cite{BP90,Kraemmer94}. The
 non-static internal propagators need not  be resummed since
 the corresponding Matsubara frequencies ensure an infrared
cut-off of the order of $T$, relative to which all the  thermal corrections
are perturbative. This relative simplicity
of resummation for static quantities allows different
approaches to the detailed calculations. At low orders one can
in general perform the resummation of diagrams by
``inspection''\cite{inspect}. In static QED calculations,
a ``resummation by inspection'' (let us call this method (a))
is even feasible at very high
orders \cite{CoPa} because the fermion lines are always hard in imaginary
time and so do not require dressing. On the other hand, in
theories with self-interacting bosonic fields more efficient and
practical  methods of higher order calculations are required
which perform the equivalent resummation, of which
there are at least three: (b) truncation of the full skeleton
expansion\cite{Akhiezer60,staticres};
(c) the use of rearranged lagrangians incoporating the screening
masses\cite{SW,BP90,RP92,AE}
and, (d) the use of dimensionally  reduced effective lagrangians
obtained by systematically integrating out of the
heavy modes \cite{Nadkarni83,Braaten94} (and references therein).
In this paper we will employ  resummation by inspection,
method (a), for the QED calculations and the truncation of
Schwinger-Dyson equations (method (b) above) for SQED. By way of
comparison, we also discuss the calculation in SQED using the
effective lagrangian, method (d). Method (c)will not be used
in this paper but a recent discussion at high orders may be found
in Ref.\cite{PS}.

As a result of the resummation, the
perturbative expansion
for the electric mass involves {\it odd} powers of the coupling strength,
that is, it is not analytic with respect to $e^2$.
This becomes apparent in spinor QED
only at the order $e^5$, but  is already manifest at the order
$e^3$ in scalar QED\cite{Kalashnikov80}, as well as in QCD\cite{Rebhan93}.
Odd powers of the coupling occur in the
perturbative expansion since, after dressing the soft propagators with the
corresponding screening masses,
 the relevant expansion parameter in the infrared is $e^2(T/m) \sim e$,
rather then $e^2$.
 From the point of view of its infrared behaviour, SQED is more
interesting than spinor QED since it involves interacting bosonic
fields. As a consequence, the scalar theory bears more ressemblence to
QCD, by exhibiting some of the non-trivial IR structure of this latter
theory, but in the (technically less involved) context of an abelian
gauge structure. In this respect, SQED serves as a toy model for
investigating resummations techniques
 to be eventually applied in high-temperature QCD (see,
e.g., Refs. \cite{Arnold94} for such recent applications).

The plan for the rest of the paper is as follows.
In Sect. 2 we display our notation and conventions, and
derive the relations (\ref{pot}),  (\ref{derpress})
and (\ref{mag}) in an exact, but formal, manner. Although most
of the material in this section may be found scattered in the
literature, we have
collected it in one place to keep the discussion self-contained.
In Sect. 3 we study massless QED in perturbation theory.
The highlights in this section are the
direct calculation of $m_D^2$ to order $e^5T^2$, discussion of
its gauge-invariance, and an all-order proof for the vanishing
of the static magnetic screening mass.
Section 4 is devoted to scalar QED. The notable results
obtained here are
$\Pi_{\mu\nu}(0, p\to 0)$ and $m_D^2$ to order $e^4 T^2$, and the
vanishing of the magnetic screening mass to the same order.
In Sect. 4.6, we re-discuss the results for SQED from the point of view
of the effective three-dimensional theory for static fields. This
sheds a new light on the resummation, and  helps keeping track of the
various terms in the diagramatic expansion.  We conclude in Sect. 5 with a
summary of results and some discussion. Some technical details omitted from
Sect. 4 are collected in the Appendices.

\setcounter{equation}{0}
\section{Notation and General Results}

\subsection{Conventions}

We summarise here our conventions and notation.
Unless otherwise stated, all
calculations from Sect. 3 onwards will
be for the massless theories at zero
chemical potential.
Ultraviolet divergences are regulated by dimensional
 continuation ($4 \to D=4-2\epsilon$)
 and renormalisation is  via minimal subtraction.
We employ the  imaginary time formalism and
 denote the four-momenta by capitals,
 $Q_\mu = (q_0,{\nbf q})$, $q_0 = i\omega_n
= i n\pi T$, with $n$ even (odd) for bosonic (fermionic) fields.
The scalar product is defined with a Minkowski metric, so that
$Q^2=q_0^2 - {\nbf{q}}^2$.  The  measure of loop integrals will
be denoted by the following condensed notation:

\beq\label{mesure}
\int[{\nrm d}Q]\equiv T\sum_{n, even}\int ({\nrm d} {\nbf q})\, ,
\qquad
 \int\{{\nrm d}Q\}\equiv  T\sum_{n, odd}\int ({\nrm d} {\nbf q})\, ,
\qquad \int[{\nrm d}Q]^\prime \equiv T\sum_{n\ne 0,
 even}\int ({\nrm d} {\nbf q})\, ,
\nonumber\eeq
where
\beq
\int ({\nrm d} {\nbf q})
\equiv \int\frac{{\nrm d}^{D-1}q}{(2\pi)^{D-1}}.\nonumber \eeq
Note that in Sect.3 we keep the fermions as four-component objects,
${\nrm Tr} (\gamma_{\mu}\gamma_{\nu})=4 g_{\mu\nu}$ eventhough
the rest of the lagrangian is dimensionally continued.

To simplify the reading of the forthcoming sections, we
 recall here the tensorial structure
of the photon propagator $D^{\mu\nu}$
($D_{\mu\nu}^{-1}=D_{0,\mu\nu}^{-1}+\Pi_{\mu\nu}$)
in the general case, and also in the hard thermal loop approximation. We follow
 closely the conventions of Ref. \cite{Weldon}, and write
\beq
\Pi^{\mu\nu}(P)={\cal P}_L^{\mu\nu}\Pi_L(p_0,{p})+ {\cal
P}_T^{\mu\nu}\Pi_T(p_0,{p}),
\eeq where $P^\mu=(p_0,{\nbf p})$, $p=|{\nbf p}|$, $\hat p^i=p^i/p$, and
the subscripts $L$ and $T$ refer to longitudinal and transverse directions
with respect to the vector ${\nbf p}$:
\beq
{\cal P}_T^{00}={\cal P}_T^{0i}=0\qquad\qquad {\cal
P}_T^{ij}=\delta^{ij}-\hat
p^i\hat p^j  \nonumber \\
{\cal P}_L^{\mu\nu}+{\cal P}_T^{\mu\nu}=\frac{P^\mu
P^\nu}{P^2}-g^{\mu\nu}.
\eeq In the static limit ($p_0=0$) the only non trivial components of
 $D_{\mu\nu}$ are
\beq\label{Dstat} D_{00}(0,{\nbf p})=-\frac{1}{{\nbf
p}^2+\Pi_L(0,p)},\qquad\qquad D_{ij}(0,{\nbf p})=\frac{\delta_{ij}-\hat
p_i\hat p_j}{{\nbf p}^2+\Pi_T(0,p)}+ \alpha \,\frac{\hat p_i \hat p_j
}{{\nbf p}^2}, \eeq
in the covariant gauge with gauge fixing parameter $\alpha$.
 Note that $\Pi_L(0,p)=-\Pi_{00}(0,p)$ and
$\Pi_T(0,p)=(1/2)\Pi_{ii}(0,p)$. { \it Please note that for simplicity
 most of our explicit calculations in Sects. 3 and 4
are in Feynman's gauge $\alpha =1$,
but the gauge-independence of our main results (the propagator
poles)  will be demonstrated.}

The one-loop polarization tensor for ultrarelativistic gauge plasmas has been
computed in Refs.~\cite{Fradkin65,Weldon}. The  dominant contribution in
the high-temperature limit (i.e. for external momenta which are small
compared to the temperature; e.g. $p_0$ and $p$ of order $eT$) is
the hard thermal loop, which has the same structure
 for both abelian and non-abelian plasmas\cite{Weldon,BP90,FT90}:
\beq \label{Pihtl}
\Pi_{\mu\nu}^{(2)} (P) =m^2\left\{ -g_{\mu 0}g_{\nu 0}+
p_0\int \frac{{\nrm d}\Omega}{4\pi}\,\frac{v_\mu v_\nu}{p_0-{\nbf
v}\cdot{\nbf p}}\right\},
\eeq
where $m^2 = e^2T^2/3$ (with $e^2\to g^2 N$ for a SU($N$)
gauge plasma), $v_\mu\equiv (1,{\nbf v})$,
$|{\nbf v}|=1$ and the angular integral $\int {\nrm d}\Omega$ runs
over all the directions of  ${\nbf v}$.
We have indicated the order of perturbation theory as a superscript on the
polarisation tensor, a convention that we shall systematically use througout.
The structure (\ref{Pihtl}) has a classical origin, as shown by thesimple
kinetic derivation\cite{BI} which is briefly
reviewed in Appendix A.
For soft $P$, $D_0^{-1}\sim P^2  \sim e^2T^2 \sim \Pi_{\mu\nu}^{(2)}$,
and the hard thermal loop must be included in   the photon propagator. We
denote this by ${}^*D_{\mu\nu}$:
 ${}^*D_{\mu\nu}^{-1} = D_{0,\mu\nu}^{-1} + \Pi_{\mu\nu}^{(2)}$.
In the static limit,
\beq\label{HTL}
\Pi_{00}^{(2)}(0,{\nbf p})=-m^2\qquad\qquad
\Pi_{ii}^{(2)}(0,{\nbf p})=0,
\eeq
so that
\beq\label{D*}
{}^*D_{00}(0,{\nbf p})=-\frac{1}{{\nbf p}^2+m^2}\qquad\qquad
{}^*D_{ij}(0,{\nbf p})=\frac{\delta_{ij}}{{\nbf p}^2}\,.
\eeq

\subsection{Exact relations}

To derive eq.~(\ref{pot}), consider the plasma in the presence of weak
static external sources with charge density $\rho^{ext}({\nbf x})$. The
free energy in
the presence of the sources is $F=-(1/\beta)\ln {\nrm Tr}\exp\{
-\beta(H+H_1)\}$, where $H_1$ is  the Hamiltonian describing the interaction
between the gauge fields and the external sources: $H_1=\int {\nrm d}^3
x\,\rho^{ext}({\nbf x})\,A_0 ({\nbf x})$. We assume that the average values
of
the gauge fields vanish in equilibrium, i.e. in the absence of sources.
 Then, to
second order in
$\rho^{ext}$, the modification in the free energy reads
\beq\label{linear}
F&=&F_0\,-\,\frac{1}{2}\int ({\nrm d}{\nbf p})\,\rho^{ext}({\nbf p})\,
D_{00}(0, {\nbf p})\,\rho^{ext}({\nbf -p})\\ \nonumber
&=&F_0\,+\,\frac{1}{2}\int ({\nrm d}{\nbf p})\,\frac{\rho^{ext}({\nbf p})\,
\rho^{ext}({\nbf -p})}{{\nbf p}^2 + \Pi_L(0, {\nbf p})}\,,
\eeq
where $F_0$ is the free energy in the absence of sources, and $D_{00}(0, {\nbf
p})$ is the exact electrostatic propagator in equilibrium. By choosing
$\rho^{ext}({\nbf x}) = q_1\delta ({\nbf x - x}_1) + q_2\delta ({\nbf x -
x}_2)$, one extracts from the above equation the interaction energy of two
isolated charges in the medium (${\nbf r}= {\nbf x}_1 - {\nbf x}_2$):
\beq
V(r) = q_1 q_2 \int ({\nrm d}{\nbf p})\,{e^{i \vec{p} \cdot \vec{r}}
\over p^2 + \Pi_L(0,p) } \,,\eeq
which is eq.~(\ref{pot}).

One way to understand eq.(\ref{linear}) is to recall that the free energy
$F[\rho^{ext}]$ is the generating functional of connected Green's
functions. Alternatively, one can express the free energy in
terms of the average gauge field
$A_0$. This is achieved by performing
the Legendre tansform
 $F'[A_0]\,=\,F[\rho^{ext}]\,-\,\int {\nrm d}^3x\,\rho^{ext}\,A_0$. In
 $F'$, the
term quadratic in $A_0$ involves the inverse propagator:
\beq
F'&=&F_0+\frac{1}{2}\int ({\nrm d}{\nbf p})\,A_0({\nbf p})\,
D_{00}^{-1}(0, {\nbf p})\,A_0({\nbf -p}) \,+\, ...\\ \nonumber
&=&\,-\,\frac{1}{2}\int ({\nrm d}{\nbf p})\,A_0({\nbf p})\,
({\nbf p}^2 + \Pi_L(0, {\nbf p}))\,A_0({\nbf -p}) \,+\, ...\,\,.\eeq
Thus
\beq\label{pres}
\Pi_L(0,0)\,=\,-\frac{1}{V}\,\frac{\delta^2 F'}{\delta A_0^2({\nbf p}=0)=
}
\bigg |_{A_0=0}\,,\eeq
where $V$ is the volume of the  plasma.
We note now that the chemical
potential $\mu$ enters the calculation  of the partition function the same w=
ay
as $A_0$ does, that is, it amounts to adding to the Hamiltonian a term
$\,-\,\mu \int {\nrm d}^3 x\,\rho_e ({\nbf x})\,=\,-\,\mu\,\rho_e({\nbf p}
 =0)$,
where $\rho_e({\nbf x})$ is the charge density operator
(in QED, $\rho_e({\nbf x}) = \psi^\dagger ({\nbf x}) \psi ({\nbf x})$).
Thus, a change
of  $A_0({\nbf p}= 0)$ is equivalent to a change of $\,-\,\mu/e$.
Since $F'$ and the pressure are related by the thermodynamic
relation $F'\,=\,-PV$, eq.~(\ref{pres}) is the same as
eq.~(\ref{derpress}).

In order to establish eq.~(\ref{mag}), we rely on the exact
 Dyson-Schwinger equations and on the Ward identities. This will also
provide us with an alternative proof of eq.~(\ref{derpress}).
 Notice first that
gauge symmetry ensures the transversality of the
polarisation tensor,
\beq
P^{\mu} \Pi_{\mu \nu}(p_0,p) = 0 \, ,
\eeq
so that
\beq
p^{i} \Pi_{i \nu}(0,p) =0 \, ,
\eeq
which implies $\Pi_{i 0}(0,p) =0$ and the transversality of
$\Pi_{ij}(0,p)$ in the spatial indices.

For spinor QED, the relevant Dyson-Schwinger equation reads
\beq
\Pi_{\mu\nu}(p_0,p) = - e^2 \int \{{\nrm d} K\} \ {\nrm Tr}
 \Bigl( \gamma_{\mu} S(P+K) \Gamma_{\nu}(P+K,K)
 S(K) \Bigr) \, , \label{sd} \eeq
where $S$ is the full fermion propagator, $\Gamma$ the full vertex, and the
trace is over spinor indices. In the limit $p_0=0,p \to 0$, eq.~(\ref{sd})
combined with the Ward identity $\Gamma_{\nu}(K,K) = { \partial S^{-1}(K)
/ \partial K^{\nu} }$ gives
\beq\label{WI}
\Pi_{\mu\nu}(p_0=0,p\to 0) = - e^2\, {\nrm Tr}\,
\gamma_{\mu} \int \{{\nrm d}K\}\, { \partial
S(K) \over \partial K^{\nu} } \, . \eeq
In the imaginary-time formalism, the variable
 $k^0 = i\omega_n +\mu_e$ takes  discrete values only.
Then, the  derivative with respect to $k^0$ in the above equation
is meant to be done on the analytic continuation of $S(k_0)$.

In the calculation of $\Pi_{(00)}$ from the r.h.s. of  eq.~(\ref{WI}),
one can replace $\del/ \del k_0 \to \del/ \del \mu_e$;
 then, by noticing  that
\beq {\nrm Tr}\,
\gamma_{\mu} \int \{{\nrm d}K\}\, S(K) \,=\, j_\mu \,= \,g_{\mu 0} \,
\rho_e\eeq
 is the charge density in equilibrium, and that $\rho_e =
\del P/ \del \mu_e$, we recover the identity (\ref{derpress}).
As for the $(i,j)$ components, they vanish after an integration by
parts, in agreement with eq.~(\ref{mag}).

A similar discussion may be carried on for SQED.
The corresponding Dyson-Schwinger equation is illustrated in Fig. 1.
To save writing, we consider directly the limit $p_0=0,p \to 0$ in these
diagrams, and denote $\Pi_{\mu\nu}\equiv \Pi_{\mu\nu}(p_0=0,p \to 0)$
in the remaining of this section.
 It is convenient to combine the two diagrams in Figs. 1.a and 1.b
by writing
\beq\label{SDab}
\Pi_{\mu\nu}^a + \Pi_{\mu\nu}^b=
-2g_{\mu\nu}\,e^2\int[{\nrm d}K]\,S(K)
-2e^2\int[{\nrm d}K]\,K_\mu\,\Gamma_\nu(K,K)\,S^2(K)
\eeq
where $S(K)=1/(-K^2+\Sigma(K))$ is the exact scalar propagator,
and $\Gamma_\nu(K,Q)$ is the vertex with one photon and two
scalar external lines, with $K$ ($Q$) denoting
 the momentum carried by theincoming (outgoing) charged particle.
We use the same Ward identity as in spinor QED to rewrite eq.~(\ref{SDab}) as
\beq\label{WI1}
\Pi_{\mu\nu}^a + \Pi_{\mu\nu}^b=
-2e^2\int[{\nrm d}K]\,{\del \over \del K^{\nu} }
\Bigl( K_\mu S(K)\Bigr) \, . \eeq
After an integration by parts
(with the assumption that $k_i S(K) \to 0$ as $|{\nbf k}| \to \infty$),
we obtain $\Pi_{ij}^a + \Pi_{ij}^b =0$.
As for the electric piece, this is rewritten by introducing a small
 chemical potential for the charged particles, so that  $k^0 \equiv
i\omega_n +\mu_e$ and
\beq\label{SDe1}
\Pi_{00}^a + \Pi_{00}^b=
-2e^2\,\frac{\del}{\del \mu_e}\,\int[{\nrm d}K]\,\Bigl(k_0 S(K)\Bigr) \, .
\eeq
The contribution of the remaining diagrams, Figs. 1.c and 1.d, is evaluated
as
\beq\label{SDcd}
\Pi_{\mu\nu}^c + \Pi_{\mu\nu}^d \,=\,
2e^2\int[{\nrm d}K] \int[{\nrm d}Q]\,
\biggl \{g^{\sigma\rho} S(K)\,D_{\mu\sigma}(K-Q)\,S(Q)\,\Gamma_{\rho\nu}
(K-Q,0,K,Q)\\ \nonumber
-\,2 S^2(K)\,\Gamma_{\nu}(K,K)\,S(Q)\,
D_{\mu\sigma}(K-Q)\,\Gamma^\sigma(Q,K)\biggr\}
\eeq
where $\Gamma_{\mu\nu}$ is the vertex
between two photons and two charged scalars. This vertex satisfies
\beq\Gamma_{\rho\nu}(K-Q,0,K,Q)\,= -e \,\left (\frac{\del }{\del K^\nu}
+\frac{\del }{\del Q^\nu}\right )\,\Gamma_\rho(K,Q)\,.\eeq
By using the Ward identities above, one obtains
 $\Pi_{ij}^c + \Pi_{ij}^d =0$ and
\beq\label{SDe2}
\Pi_{00}^c + \Pi_{00}^d=
-2e^3\,\frac{\del}{\del \mu_e}\,\int[{\nrm d}K] \int[{\nrm d}Q]
\,\Bigl(S(K)\,S(Q)\,D_{0\rho}(K-Q)\,\Gamma^\rho(K,Q)\Bigr) \, . \eeq
 We recognize in the r.h.s.  of eqs.~(\ref{SDe1}) and (\ref{SDe2})
 the average electric charge density
$\rho_e \,=\, i(\phi^\dagger \del_0 \phi - (\del_0 \phi^\dagger)\phi)
\,-\,2e\,A_0\phi^\dagger \phi$ expressed in terms of exact propagators
and vertices. Thus
\beq\label{SDe}
\Pi_{00}\equiv \Pi_{00}^a + \Pi_{00}^b+\Pi_{00}^c + \Pi_{00}^d \,=\,
- e^2 \frac{\del \rho_e}{\del \mu_e}\,=\,-e^2\,\frac{\del^2 P}{\del \mu_e^2}\,.
\eeq
It is interesting to observe that the vanishing of $\Pi_{ij}$
is obtained by independent  cancellations
of the diagrams 1.a and 1.b among themselves, and of the diagrams 1.c and 1.d
among themselves.

Note that the  manipulations above are formal in the sense
that we have left aside the question of the
regularization of UV divergences, as well as possible IR problems.
We will not address these questions in general here but
in the following sections we shall go through an
explicit perturbative verification.

\setcounter{equation}{0}

\section{Perturbative QED }

\subsection{ The static-infrared limit}
We now begin our explicit calculations by considering in this
subsection the object $\Pi_{\mu \nu}(0,p \to 0)$ up to fifth order
($e^5$). The explicit computations here have been performed
in the Feynman gauge. At one loop, a standard calculation gives
\beq
\Pi_{\mu\nu}^{(2)}(0,p\to0) &=& (e\mu^{\epsilon})^2
 \int \{dK\} \ {{\nrm Tr} \
(\gamma_{\mu} {\slashchar K} \gamma_{\nu}{\slashchar K}
) \over K^4} \nonumber \\ &=&
-4 (e\mu^{\epsilon})^2
\int\{dK\} {g_{\mu\nu}K^2 -2 K_{\mu}K_{\nu} \over K^4} \, , \label{1lp}
\eeq with ${\slashchar K}\equiv \gamma_\rho K^\rho$.
As usual, $\mu$ is the mass scale introduced by dimensional regularisation,
and the gauge coupling $e$ is dimensionless.
 Within dimensional regularisation, an integration by
parts in the ${\nbf{k}}$ integral gives \beq
\int \{dK\} {{\nbf k}^2 \over K^4} = {-(D-1) \over
 2} \int { \{dK\} \over K^2} \, . \label{ibp}
\eeq
Hence one deduces from (\ref{1lp}),
\beq\label{mag2}
\Pi_{ij}^{(2)}(0,p\to0) = 0 \; ,
\eeq
and
\beq\label{el2}
m^2 \equiv -\,\Pi_{00}^{(2)}(0, p\to 0)
&=& -4 (e\mu^{\epsilon})^2 (2-D) \int { \{dK\} \over K^2} \\
&=&{e^2T^2 \over 3} \, ,
\eeq  where the limit $D\to 4$ has been taken in the second line.
In the rest of this paper, this limit will often be taken
implicitly in our final results.
The fermionic integral occurring above is a special case of \beq
f_n &\equiv& \int { \{d\,Q\} \over (Q^2)^n} = (2^{2n+1-D}-1) \ b_n \, ,
\eeq
where
\beq\label{bn}
b_n &\equiv& \int {[d\,Q]\over (Q^2)^n} \, \nonumber \\ && \nonumber \\
&=& T^{D-2n}\, {2 \ (-1)^n \ \pi^{{D-1 \over 2}}
 \over (2 \pi)^{ 2n} \ \Gamma(n) }
\, \zeta(2n +1 -D) \ \Gamma\left({2n+1-D \over 2}\right) \, ,
\eeq
and $\zeta(x)$ is Riemann's {\it zeta}-function.
The quickest way to evaluate the $b_n$ is to first
integrate over the momenta and then perform the frequency sum.
The integral $f_n$ is then obtained using the following trick:
consider the sum $ b_n + f_n$ and rescale the momenta there by a
factor of $2$, giving $b_n + f_n = 2^{2n+1-D}b_n$, from which
the quoted result follows.
Note that, because of the dimensional continuation, the Matsubara mode
$q_0 = 0$ does not contribute to the integral in eq.~(\ref{bn}).

The next contribution is of order $e^4$ and comes from the two loop
diagrams shown in Fig. 2.
Each of these diagrams is quite complicated, even in the static IR limit, but
their sum is remarkably simple and is given by \beq
\Pi_{\mu\nu}^{(4)}(0,p\to0) = 4(e \mu^\epsilon)^{4} (D-2) (
\Pi_{\mu\nu}^{(4o)} + \Pi_{\mu\nu}^{(4e)} )
\eeq
with
\beq
\Pi_{\mu\nu}^{(4o)} &=& \int \{dK dR\} {K_{\mu}R_{\nu} + K_{\nu}R_{\mu} \over
K^4 R^4 } \, , \label{4o} \\
\Pi_{\mu\nu}^{(4e)} &=& (b_1 -f_1) \int \{dK\} \left({g_{\mu\nu} \over
 K^4} -
{4K_{\nu} K_{\mu} \over K^6} \right) \; . \label{4e} \eeq
Note that eq.~(\ref{4o}) vanishes at zero chemical potential (at
nonzero chemical potential it only contributes to $\Pi_{00}$).
 Using integration by parts, as in (\ref{ibp}), one obtains \beq\label{mag4}
\Pi_{ij}^{(4)}(0,p\to0) =0 \, ,
\eeq
and
\beq
\Pi_{00}^{(4)}(0,p\to0) \,=\, 4 (e\mu^{\epsilon})^4\, (D-2)(D-4)(b_1-f_1)f_2
\,=\, {e^4 T^2 \over 8 \pi^2}\, . \label{el4} \eeq
Notice the cancellation between the UV divergencies of eqs.~(\ref{4o})
and (\ref{4e}) in their sum (\ref{el4}) (the quantity $(D-4) f_2$ being
finite for $D\to 4$). This cancellation is necessary since the
contribution from the $UV$ counterterm diagrams mutually cancel
(see later).

Consider now the behaviour of the magnetostatic polarization tensor for
small, but non-vanishing, momenta. We shall argue here that
$\Pi_{ij}(0, p\to 0) ={\cal O}(p^2)$ to all orders in perturbation theory.
To see this, consider an arbitrary graph
contributing to $\Pi_{ij}(0,p)$. Label the photon lines in the graph by a
  set of independent loop momenta so that the external momentum ${\nbf p}$
only flows along the fermion lines. Since in imaginary time the fermionic
propagators are infrared safe, they can be expanded
 with respect to the soft momentum ${\nbf{p}}$. Then the external
momentum appears only in the numerator and  rotational symmetry ensures
 that the terms with  odd powers of ${\nbf{p}}$ vanish.
Hence,  $\Pi_{ij}(0,p)$ is analytic in $p^2$ for small $p$,
so that  $\Pi_{ij}(0,p\to 0)= {\cal O}(1) +{\cal O}(p^2)$. By also
using eq.~(\ref{mag}), we conclude that  $\Pi_{ij}(0,p)$ vanishes
at least as $p^2$ as $p\to 0$.

The proof in the last paragraph is rigorous  only when applied
to diagrams without
photon self-energy insertions since these latter
cause power-like infrared
divergences along internal lines
due to the nonvanishing of $\Pi_{00}(0,0)$, thus
making the arguments formal. For diagrams with self-energy
insertions, the arguments of the last paragraph must be applied
not to individual diagrams but to sums of similar diagrams
which result in a new effective graph with dressed photon
propagators. The conclusion then is as before, namely
$\Pi_{ij}(0,p \to 0) = {\cal O}(p^2)$.
Finally, note that the proof is independent of the fermion mass.

Beyond fourth order we begin to get contributions nonanalytic in $e^2$. The
$e^5$ term comes from dressing the zero mode of the photon propagators,
as shown  in Figs. 3 (cf. \cite{CoPa}).
The nonzero modes of the photon line (like the modes
of the fermion lines) are cut off in the infrared by the scale $T$ and
dressing those modes just gives the usual perturbative corrections
(which are analytic with respect to $e^2$). Consider now
a static  internal photon line:
 inserting the {\it electric} polarisation tensor along this line
 causes infrared divergences which can be summed up,
with the result that the bare electrostatic  propagator gets replaced by
${}^*D_{00}(0,{\nbf q})=-{1}/({\nbf q}^2+m^2)$,
$m^2=e^2 T^2/3$ (recall eq.~(\ref{D*})).
On the other hand, insertions of the static {\it magnetic}  polarisation
 tensor give only perturbative corrections,
 as one can verify from power counting by using $\Pi_{ij}^{(2)}(0, {\nbf q})
={\cal O}(q^2)$ (also recall that $\Pi_{0i}(0,q)=0$ to all
orders so that there are no corresponding insertions).

The fifth order contribution is then obtained from the diagrams
in Fig. 3, in which the internal photon line is static and is associated
with the dressed propagator  ${}^*D_{\mu\nu}(0,{\nbf q})$,
eq.~(\ref{D*}).  It is then convenient to write
\beq\label{*d}
{}^*D_{\mu\nu}(0,{\nbf q}) &=& D_{0,\mu\nu}(0,{\nbf q}) + { m^2 g_{\mu0}
 g_{\nu0} \over {\nbf q}^2({\nbf q}^2+m^2)} \nonumber \\
&\equiv&  D_{0,\mu\nu}(0,{\nbf q}) + {}^*d_{\mu\nu}(0,{\nbf q}),\eeq
and to  observe that it is the piece ${}^*d$  which is responsible for the
$e^5$ contribution.  In fact, to order $e^5$, a further approximation
can be done. This consists in neglecting  the $q$
dependence along the fermion lines.  Then, the $q$ integral
decouples and is easily evaluated: it is of order $\int
{{\nrm d}^3q\, m^2 / q^2(q^2+m^2)} \sim m\sim e$, and,
since one has a factor of $e^4$ from the four vertices, the net
result is  of order $e^5$. The forgotten $q$-dependence along the fermion lines
only gives a subleading contribution because the fermion lines are IR safe,
so that one may expand out the $q$ dependence and do the usual power
counting.

Thus the exact fifth order contribution comes from the sum of the diagrams
in Fig. 3 when
${}^*d(Q)$ is used for the photon lines and the $Q$ dependence along the
fermion lines is ignored. We get:
\beq
\Pi_{\mu\nu}^{(5)}(0,0) = -4 (e\mu^{\epsilon})^4
 \,m^{D-3}T \int{ (dx) \over x^2
(x^2+1)} \int {\{dK\} \over K^8} S_{\mu\nu} \, , \eeq
where
\beq
S_{\mu\nu} &=& (4K_{\mu} K_{\nu} -2 g_{\mu\nu}K^2 )(K^2+2k^2) +
K^4(2g_{\mu0}g_{\nu0} - g_{\mu\nu}) \nonumber \\ && \;\;\;\; - 8k_0 K^2
(g_{\mu 0}K_{\nu} +g_{\nu 0}K_{\mu}) + 16K_{\mu}K_{\nu}(K^2+k^2) \, .
\eeq
Again using manipulations as in (\ref{ibp}) gives \beq
\Pi_{ij}^{(5)}(0,p\to0) =0 \, ,
\eeq
and
\beq\label{el5}
\Pi_{00}^{(5)}(0,p\to0) &=& 4 (e\mu^{\epsilon})^4
 \,m^{D-3} T\, (D-2)(D-4) f_{2}
\int{ (dx) \over x^2(x^2+1) } \nonumber \\ &=& {- e^5 T^2 \over 4 \pi^3
\sqrt{3}} \, . \eeq
In summary we have shown for massless QED in the Feynman gauge at
temperature $T$ and zero chemical potential $\mu_e$, \beq
\Pi_{ij}(0,p\to0) ={\cal O}(p^2) + {\cal O}(e^6T^2) \, . \label{magans} \eeq
and
\beq
\Pi_{00}(0,p\to0) = - T^2 \left( {e^2 \over 3} - {e^4 \over 8\pi^2} + {e^5
\over 4\pi^3 \sqrt{3}} \right) + {\cal O}(e^6 T^2) \, . \label{elans} \eeq
A scan of the computations shows that
Eq.(\ref{magans}) has also been verified for nonzero chemical potential.
The result (\ref{elans}) --- which is the left-hand-side of the identity
eq.~(\ref{derpress}) --- is in accordance with the r.h.s. of as given in
\cite{Kapusta89}. Also, since the right-hand-side of (\ref{derpress})
is gauge-independent, this implies the gauge-independence of
(\ref{elans}) though we have done our calculations in the
simpler Feynman gauge.

The result eq.~(\ref{elans}) has turned out to be  UV finite
eventhough we seem to have ignored the UV renormalisation.
However recall first that zero-temperature  counterterms suffice to
render the theory UV finite, and that gauge invariance ensures that the
vacuum (i.e. zero-temperature)  polarization tensor vanishes in the
zero-momentum limit. Then, in our calculation, no
counterterms were needed for the subdiagrams of the two loop
polarization tensor for the following reasons:
i) The order $e^4$ vertex and fermion
wave-function counterterm diagrams cancel against each other
because of the Ward identity $Z_1 =Z_2$; ii) since we are working
in the massless limit, there is no mass counterterm;
iii) there is no order $e^5$ counterterm diagram.
On the other hand, UV divergences will
make their appearance when the polarization tensor will be considered
for non-vanishing momenta in the next section.
Similar arguments apply to the SQED calculation in Sect.4,
and so we shall not repeat them there.

\subsection{Screening masses}

Having computed $\Pi_{\mu \nu}(0,p \to 0)$, we are ready to
determine the screening masses.
To one-loop order, the electrostatic propagator ${}^*D_{00}(0,{\nbf
p})=-{1}/({\nbf p}^2+m^2)$ has a simple pole in the upper half
of the complex $p$-plane, occuring at $p=im$. Accordingly, $m_D^2 = m^2
+{\cal O}(e^4T^2)$ (recall eq.(\ref{debyemass})).
The higher-order corrections that we have calculated
 modify the position
of this pole, without changing the analytic structure of $D_{00}(0,{\nbf p})$.
Accordingly, we may define the electric screening mass
as the solution of the equation \beq
p^2 - \Pi_{00}(0,p) =0 \label{rt}
\eeq
for $p \sim eT$. Since we will only do the calculations up to order $e^5$,
we need the expansion of the one and two loop static polarisation tensors
up to the following orders, \beq\label{002}
\Pi_{00}^{(2)}(0,p \sim eT) &=& -\, e^2T^2\left( a_{20} +
{a_{21} \ p \over T} + {a_{22} \ p^2 \over T^2} +
{a_{32} \ p^3 \over T^3} \right) + {\cal O}(e^2p^4/T^2) \\
\Pi_{00}^{(4)}(0, p \sim eT) &=& e^4 T^2 \left(a_{40} + {a_{41}
\ p \over T}\right)
+ {\cal O}(e^4p^2) \, ,
\eeq
together with $\Pi_{00}^{(5)}(0,0)$.

{}From the last subsection we have $a_{20} = 1/3$ and $a_{40} =1/8\pi^2$.
Furthermore $a_{21}=a_{23}=a_{41}=0$
for the reasons explained after eq.~(\ref{el4}).
We now calculate $a_{22}$ starting from \beq
\Pi_{00}^{(2)}(0,p) = 2(e\mu^{\epsilon})^2
 \int \{dK\} \delta_{p_{0},0} \left( {2 \over
K^2} + { 4k^2- p^2 \over K^2(K+P)^2} \right) \, .\label{q1}
\eeq
Since $k_0$ is discrete and of order $T$, one may expand \beq
{ \delta_{p_{0},0} \over (K+P)^2 } = {{ \delta_{p_{0},0}
 \over K^2}} \left(1 + {2{\nbf k\cdot p} + {\nbf p}^2
\over K^2} + {(2{\nbf k\cdot p} + {\nbf p}^2)^2
 \over K^4 } + ... \right) \, .\label{q2}
\eeq
Then using (\ref{q2}) in (\ref{q1}) and simplifying , we get \beq
\Pi_{00}^{(2)}(0,p) = \Pi_{00}^{(2)}(0,0)\, -\, { 2 (e\mu^{\epsilon})^2
p^2 \over 3}(D-2)f_2 + {\cal O}(e^2p^4/T^2) \, . \label{q3}
\eeq
The order $e^2p^2$ term in (\ref{q3}) is UV divergent as $D \to 4$ and this
divergence is cancelled by the photon wave-function renormalisation
counterterm $\delta Z_3 p^2$, with $\delta Z_3= - e^2/12\pi^2\epsilon$.
 Then the renormalized value of $a_{22}$ reads \beq
a_{22}^{R} = {4 \mu^{2\epsilon} (1-\epsilon)
 \over 3} f_2 - { 1\over 12\pi^2 \epsilon} \,
\,=\, \frac{1}{12\pi^2} \left(
\gamma- 1 + \ln\,{4\mu^2 \over \pi T^2} \right)\, .\eeq
We summarize the above results by writing ($p\sim eT$)
\beq
\Pi_{00}(0,p)= - m^2 - a_{22}^{R}e^2 p^2 + \Pi_{00}^{(4)}(0,0) +
\Pi_{00}^{(5)}(0,0) + {\cal O}(e^6 T^2)\, .
\eeq
The solution $p^2= -m_D^2$  of eq.~(\ref{rt}) to fifth order is therefore
\beq
m_D^2 = T^2 \left( {e^2 \over 3} - {e^4 \over 8\pi^2} -{e^4 \over 36 \pi^2}
\left[\gamma- 1 + \ln \,{4\mu^2 \over \pi T^2} \,\right]
+ {e^5 \over 4\pi^3\sqrt{3}} \right) + {\cal O}
(e^6 T^2) \, . \label{elfin} \eeq
Here, $e\equiv e(\mu)$ is the running coupling constant defined by the minimal
subtraction scheme. Since this satisfies ${\nrm d} e/{\nrm d}\ln \mu \,=\,
e^3/12 \pi^2$, it is clear that the r.h.s. of eq.~(\ref{elfin}) is
independent of the renormalization scale $\mu$, to order $e^5$.

Eq.~(\ref{elfin}) is
 our result for the electric screening mass of massless QED at
temperature $T$ and zero chemical potential. Up to order $e^4$ this
coincides with the result of Ref. \cite{Rebhan93}.
As for the magnetic screening mass, this vanishes  as expected,
since $\Pi_{ij}(0,p \to 0) ={\cal O}(p^2)$.

Let us now establish the gauge-independence of our result (\ref{elfin}).
The constants $a_{2n}$ come from the one-loop diagram and so are manifestly
gauge-independent. The constant $a_{40}$ is gauge-independent because of the
relation (\ref{derpress}) while the vanishing of $a_{41}$
is a gauge-independent
statement
 since the arguments following eq.~(\ref{el4}) make no
reference to any gauge-choice. Hence the electric screening mass (and
similarly the vanishing magnetic screening mass) to fifth order as given by
eq.~(\ref{elfin}) is gauge-independent.
\setcounter{equation}{0}
\section{Perturbative Scalar QED}

In this section we consider the electromagnetic interactions of a
charged scalar field $\phi$ described by the Lagrangian
($D_\mu=\del_\mu+ieA_\mu$)
\beq\label{SQED}
{\cal L}=-\frac{1}{4}F_{\mu\nu}^2-\frac{1}{2 \alpha}(\del\cdot A)
^2+(D_\mu\phi)^\dagger(D^\mu\phi)-
\frac{\lambda}{4}(\phi^\dagger\phi)^2.
\eeq
Since we are interested
only in the effects of the electromagnetic interactions, we shall
ignore the self coupling of the complex field $\phi$, i.e we
assume $\lambda\to 0$ in what follows.

In this section, we shall compute the correction of order $e^4T^2$
to the Debye mass, and we shall verify that $\Pi_{ij}(0, {\nbf p})
= {\cal O}(p^2)$ as $p\to 0$, to the same order.
In order to avoid double counting in higher order  calculations,
we shall refer to the skeleton diagrams  displayed in Fig. 1.
The corrections to $\Pi_{\mu\nu}$ at various orders will be obtained
by expanding to the appropriate order
the exact propagators or vertices in these  diagrams.
As explained before, in this procedure one must keep the thermal masses
on the static propagators, while  non-static propagators can be
perturbatively expanded.

\subsection{Leading order results: hard thermal loops}

In leading order, the self-energies for  the photon and the scalar particle
 are obtained from the 1-loop diagrams in Figs. 4 and 5
respectively. For soft external momenta (that is, for
 $p_0$ and $p$ of order $eT$), the  dominant contributions to these diagrams
come from loop momenta of the order of $T$: these are the hard thermal
 loops\cite{BP90,FT90}. The photon hard thermal loop
was already presented in Sect. 2
(eqs.~(\ref{Pihtl})--(\ref{D*})). The hard thermal loop for the scalar
self-energy reduces  to a (gauge-independent) local mass term:
\beq\label{Sihtl}
\Sigma^{(2)}(P) =  (D-1)\left(e\mu^\epsilon\right)^2 \int[{\nrm d}Q]
\,S_0(Q)=\frac{e^2 T^2}{4} \,\equiv \, M^2\, ,\eeq
where $S_0(Q)= -1/ Q^2$.
We define the propagator ${}^*S$ by
 ${}^*S^{-1} = S_0^{-1}+\Sigma^{(2)}$. In the static limit,
\beq\label{S*}
{}^*S(0, {\nbf p})=\frac{1}{{\nbf p}^2 +M^2}\, .\eeq

Generally, the one-loop  results cannot be trusted beyond the
hard thermal loop approximation. This is so because, beyond leading order,
soft loop momenta start to contribute and the corresponding propagators
must include the hard thermal loops.  Consider for example the small
momentum behavior of the static one-loop polarisation tensor. It is
shown in Appendix B  that
\beq\label{1Li}
\Pi_{ii}^{1L}(0,p)=\frac{1}{8}e^2pT+{\cal O}(e^2 p^2)
\eeq
(see eq.~(\ref{1Liiexp})).
A similar  behaviour is obtained for the scalar one-loop self-energy:
\beq\label{linp}
\Sigma_{1L}(0,p)= M^2 - \frac{1}{4}e^2pT+{\cal O}(e^2 p^2).
\eeq
These  non-analytic (in
$p^2$) contributions arise from the static internal
 modes. We shall verify  in the next subsection that these terms
 disappear once thermal masses are included in static internal lines.
In QED, such a problem does not occur because the corresponding one-loop
diagram for $\Pi_{\mu\nu}$ has only fermionic internal lines.
Note  that for the component $\Pi_{00}$
no resummation is needed to get the leading low momentum
behavior to one-loop order (see eq.~(\ref{1L00exp})): because of the
vector structure of the electromagnetic interaction, the mode
$n=0$ does not contribute to the Matsubara sum in the contribution
of the diagram 4.b to $\Pi_{00}(0,p)$; as for
 the  diagram 4.a, it is momentum independent.

\subsection{Next-to-leading order: ring summation}

In this subsection,
we consider the consequences of the resummation on the one-loop
diagrams of Figs. 4 and 5. We isolate the  {\it static} Matsubara mode, since
this is the only one which is concerned by the resummation, and
we  replace the bare propagators  by the propagators
 ${}^*D_{\mu\nu}$ and ${}^*S$ obtained in the hard thermal loop approximation.
We thus obtain the dressed one-loop diagrams of Figs. 6 and 7, whose
 contributions will be denoted by ${}^*\Pi_{\mu\nu}(0, {\nbf p})$
and ${}^*\Sigma(0, {\nbf p})$,  respectively.

The tadpole diagram in Fig. 6.a gives
\beq\label{ring}
{}^*\Pi_{\mu\nu}^{a}=
-2g_{\mu\nu}\left(e\mu^\epsilon\right)^2T\int({\nrm d}{\nbf
k})\left(
\frac{1}{{\nbf k}^2+M^2}-\frac{1}{{\nbf k}^2}\right)\, .
\eeq
The second term inside the brakets in eq.(\ref{ring}) substracts that
contribution of the static mode which has already
been included in the hard thermal loop, eq.~(\ref{Pihtl}).
Note that this term vanishes in dimensional regularisation so we
could have as well omitted it. However it is better to keep it
here. This will allow us to show that the final results, when
appropriate contributions of a given order are added, are both
ultraviolet and infrared finite, even in the absence of
regularisation.  Letting $D\to 4$ in eq.~(\ref{ring}), we get
\beq\label{3a}
{}^*\Pi_{\mu\nu}^{a}=\frac{g_{\mu\nu}}{2\pi}
\,e^2MT=\frac{g_{\mu\nu}}{4\pi}\,e^3T^2\, .
\eeq
The diagram in Fig. 6.b gives a non trivial contribution,
of order $e^3$, only to the magnetic piece $ \Pi_{ii}(0,p)$. As $p\to 0$,
\beq \label{3b}
{}^*\Pi_{ii}^{b}(0,0)=
-4 \left(e\mu^\epsilon\right)^2T\int({\nrm d}{\nbf
k}) \,{\nbf k}^2\left(
\frac{1}{\left({\nbf k}^2+M^2\right)^2}-\frac{1}{{\nbf k}^4}\right)
\\\nonumber = \frac{3}{2\pi}\,e^2 M T =\frac{3}{4\pi}\,e^3T^2,\qquad\qquad
\eeq
By adding together the contributions (\ref{3a}) and (\ref{3b}), we obtain the
total contribution of order $e^3$ to the
zero-momentum limit of the  polarisation tensor\cite{Kalashnikov80}:
\beq\label{total3}
\Pi_{00}^{(3)}(0, 0) =\frac{e^2MT}{2\pi}=\frac{e^3T^2}{4\pi},
\qquad\qquad \Pi_{ii} ^{(3)}(0, p\to 0)=0.
\eeq
The above correction to $\Pi_{00}(0,0)$ can be understood as a classical
correction, as discussed at the end of Appendix A.

Consider now the momentum dependence of ${}^*\Pi_{\mu\nu}(0,p)$. The
electric piece is independent of $p$, since it is entirely given
by the tadpole of Fig. 6.a (recall eq.~(\ref{3a})). For the magnetic piece,
both diagrams in Fig. 6 contribute, and give
\beq\label{3ii}
{}^*\Pi_{ii}(0, {\nbf p})=
 \left(e\mu^\epsilon\right)^2T\int({\nrm d}{\nbf
k}) \,\left\{\frac{2(D-1)}{{\nbf k}^2+M^2}-
\frac{({\nbf 2k+p})^2}{({\nbf k}^2+M^2)
\left(({\nbf k+p})^2+M^2\right)}\right\}.\eeq
The presence of the mass $M\sim eT$ in the denominators allows an expansion
of the last one  with respect to ${\nbf p}$. After integration over
 ${\nbf k}$, only the terms even in ${\nbf p}$ survive.
The small momentum expansion of (\ref{3ii}) is therefore
\beq {}^*\Pi_{ii}(0, {\nbf p})=  \frac {e^2 p^2}{12\pi}\,
\frac{T}{M}\left\{1+ {\nrm c}_1 \,(p/M)^2 +{\nrm c}_2 \,(p/M)^4 + ...\right\},
\eeq
where the ${\nrm c}_i$'s are constant coefficients.
Note that  there is no  term linear in $p$,
 contrary to the pure one-loop result of eq.~(\ref{1Lii}).
As $p\to 0$, the leading term is proportional to $e^2 (T/M) p^2\sim e p^2$.
 For $p\sim eT$, all the terms in the r.h.s. are of order $e^3 T^2$, and
the integral in (\ref{3ii}) should be computed
 exactly, with the following result:
\beq\label{iiop}
{}^*\Pi_{ii}(0, {\nbf p}) =  \frac{e^2MT}{2\pi}
\left\{\frac{4M^2+p^2}{2pM}\,{\nrm arctan}\,\frac{p}{2M}\,-1\right\}.\eeq
For small momenta $p\simle eT$, this equation
 gives the leading infrared behaviour of the
static magnetic polarization operator (the non-static modes in the
one-loop diagrams in Fig. 4 contribute  only to order $e^2p^2$). For
large momenta, $p/M \gg 1$,  ${}^*\Pi_{ii}(0, {\nbf p}) \to e^2 pT/8$,
as for the undressed one-loop result of eq.~(\ref{1Li}).

A similar discussion applies to ${}^*\Sigma(0,{\nbf p})$, the correction to
the static scalar self-energy given by the dressed one-loop diagrams of
Fig. 7 (which, we recall, involve only the internal mode with zero frequency).
 A straightforward computation gives
\beq\label{sig*}
{}^*\Sigma(0,{\nbf p}) =
\left (e\mu^\epsilon\right)^2T\int ({\nrm d}{\nbf q})\Biggl\{
 \frac{1}{{\nbf q}^2+m^2}
+\frac{1}{{\nbf q}^2+M^2}+ \frac{D-3}{{\nbf q}^2}
-\frac{2}{{\nbf q}^2}\frac{{\nbf p}^2-M^2}{({\nbf q+p})^2+M^2}\Biggr\},
\eeq in the Feynman gauge.  The $q$-integral
can be performed readily, and for $(D-1)\to 3$, one gets
\beq\label{SIG*}
{}^*\Sigma(0,{\nbf p}) = \frac {e^2MT}{4\pi}
\left\{ \frac{2(M^2-{\nbf p}^2)}{M p}
 {\nrm arctan}\frac{p}{M} \,-\,\frac{m+M}{M}\right\},\eeq
which admits the following small-momentum expansion:
\beq\label{sigexp}
{}^*\Sigma(0,{\nbf p}) = \frac {e^2MT}{4\pi}\left\{
\frac{M-m}{M} - \frac{8}{3} (p/M)^2 +{\cal O}(p^4/M^4)\right\}.
\eeq
 For high momenta, $p/M \gg 1$, we recover the linear behaviour
in $p$  as in  eq.~(\ref{linp}).
For $p\simle eT$, eq.~(\ref{SIG*}) gives the
dominant non-trivial momentum behaviour of the scalar self-energy.

\subsection{Order $e^4$: diagrams 1.a and 1.b}

Contributions of order $e^4T^2$ arise from two-loop diagrams in
which the static propagators are dressed
by thermal masses. Because of the resummation involved in the
thermal masses, parts of these two-loop diagrams have already been
included in the order $e^3$ calculation. In order to avoid double
counting, we refer to the skeleton diagrams of Fig. 1.
  We shall give details only for the tadpole diagram, Fig. 1.a:
\beq\label{e4tadpole}
\Pi_{\mu\nu}^a=-2g_{\mu\nu}\left(e\mu^\epsilon\right)^2\int[{\nrm
d}K]\,S(K)\equiv g_{\mu\nu}\Pi^a
\eeq
where $S(K)=1/(-K^2+\Sigma(K))$ is the exact scalar propagator, with
$\Sigma(K)$ the exact self-energy. In these expressions
$K=(2i\pi nT, {\nbf k})$. When $n\ne 0$, $\Sigma$ represents a
small perturbative correction and only the second term in the expansion
\beq\label{dysonscalar}
S(K)=-\frac{1}{K^2}\left( 1+\frac{\Sigma(K)}{K^2}+\cdots\right)
\eeq
is in fact needed to
evaluate (\ref{e4tadpole}) up to order $e^4$. When $n=0$, infrared
divergences render the expansion above meaningless. It is then
convenient to expand about the massive propagator ${}^*S(0,{\nbf k})$
(eq.~(\ref{S*})):
\beq\label{propexpansion}
S(0,{\nbf k})&=&\frac{1}{{\nbf k}^2+M^2+(\Sigma(0,{\nbf k})-M^2)}\\
\nonumber
 &=& \frac{1}{{\nbf k}^2 }+\left( \frac{1}{{\nbf
k}^2+M^2}-\frac{1}{{\nbf k}^2}\right)-\frac{\Sigma(0,{\nbf
k})-M^2}{({\nbf k}^2+M^2)^2}+\cdots
\eeq
When the expansions (\ref{dysonscalar}) and (\ref{propexpansion})
are used in eq.(\ref{e4tadpole}), the following result is obtained
\beq\label{fullpia}
\Pi^a&=&-2\left(e\mu^\epsilon\right)^2\int[{\nrm
d}K]\,S_0(K)\\ \nonumber
 &\,&-2\left(e\mu^\epsilon\right)^2T\int({\nrm d}{\nbf
k})\left[{}^*S(0,{\nbf k})-S_0(0,{\nbf k})\right]\\ \nonumber
&\,& +2\left(e\mu^\epsilon\right)^2\int[{\nrm
d}K]^\prime \,S^2_0(K)\,\Sigma(K)\\ \nonumber
&\,& +2\left(e\mu^\epsilon\right)^2T\int({\nrm d}{\nbf
k})\,{}^*S^2(0,{\nbf k})\left[ \Sigma(0,{\nbf k})-M^2\right]\\ \nonumber
&\,&\cdots
\eeq
where the neglected terms are, at least, of order $e^5T^2$.
The self-energy entering the r.h.s. is the one-loop self-energy,
that is, it is obtained from the diagrams in Fig. 5 or, if necessary,
from the dressed diagrams of Fig. 7 (see below).

Consider now the different terms in the right hand side of eq.~(\ref{fullpia}).
We have already evaluated the first two integrals giving
respectively the contributions of order $e^2$ and $e^3$. For the
third integral, it is enough (since $k^0 \sim T$)
 to use the {\it one-loop} expression of the
scalar self-energy $\Sigma$, i.e.
\beq\label{a41}
2\left(e\mu^\epsilon\right)^2\int[{\nrm
d}K]^\prime\,S^2_0(K)\,\Sigma_{1L}(K),\eeq
 where (see Fig. 5)
\beq\label{S1L}
\Sigma_{1L}(K)=M^2+2\left(e\mu^\epsilon\right)^2 \,K^2\int[{\nrm d}Q]
\,S_0(Q)\,S_0(K+Q), \eeq
in the Feynman gauge.

The evaluation of the last term in  eq.~(\ref{fullpia})
is more involved. It is again necessary to separate the static
($\omega_m = 0$) and non-static ($\omega_m\not = 0$) modes
in the one-loop diagrams giving $\Sigma(0,{\nbf k})$ (see Figs. 5 and 7);
here,  $\omega_m$ denotes the Matsubara frequency inside the loop.
For the non static modes, bare  propagators can be used, as in Fig. 5, and we
recover the $m\not = 0$ piece of the   one-loop self-energy
from eq.~(\ref{S1L}). The corresponding contribution to $\Pi^a$ reads
\beq\label{mixt}
2\left(e\mu^\epsilon\right)^2T\int ({\nrm
d}{\nbf k}) \,{}^*S^2(0,{\nbf k})\left[\Sigma_{1L} (0,{\nbf
k})-M^2\right]_{m\not =0}\qquad\\ \nonumber
=-4\left(e\mu^\epsilon\right)^4T\int({\nrm d}{\nbf
k})\,\frac{{\nbf k}^2}{\left({\nbf k}^2 + M^2\right)^2}
\int [{\nrm d}Q]^\prime \frac{1}{\omega_m^2 + {\nbf q}^2}\,
\frac{1}{\omega_m^2 + ({\nbf q}+{\nbf k})^2}.\eeq
It is not difficult to see that the above contribution is of order
$e^4 MT\sim e^5 T^2$ (in $D=4$) and therefore can be ignored in
our present computation of the order $e^4$. In fact, eq.~(\ref{mixt})
 is precisely of the type already
encountered in Sect. 3.1, in computing the contribution of order $e^5$
to $\Pi_{\mu\nu}(0,0)$ for spinor QED. In particular, its
ultraviolet singularity in the limit $D\to 4$ is harmless, since
it will be compensated by similar contributions arising from other
{\it mixed} two-loop diagrams (as happens, e.g., in eq.~(\ref{el5})).
(We characterize as ``mixed'' any two-loop graph where one of the internal
frequencies is non-vanishing, while the other one is zero.)
 Mixed graphs do not contribute to $\Pi_{\mu\nu}(0,p)$  to order $e^4$,
and will be neglected in what follows. (This applies, in particular,
to the mixed graph included in eq.~(\ref{a41}).)

Consider now the remaining contribution to eq.~(\ref{fullpia}), that is,
\beq\label{a42}
2\left(e\mu^\epsilon\right)^2T\int ({\nrm
d}{\nbf k}) \,{}^*S^2(0,{\nbf k})\left[{}^*\Sigma (0,{\nbf
k})-M^2\right]_{m=0}.
\eeq
This corresponds to two-loop diagrams
where {\it both} the internal frequencies are zero, so that
the corresponding propagators are dressed by the hard thermal loops.
The integral (\ref{a42}) is explicitly written in  eq.~(\ref{A3}) below.

We turn now to the diagram 1.b, which gives
\beq\label{fullb}
\Pi^b_{\mu\nu}(0)=-2\left(e\mu^\epsilon\right)\int[{\nrm
d}K]\,k_\mu\, \Gamma_\nu(K,K)\, S^2(K)\,.
\eeq
 To one-loop order, and for $Q=K$, one readily gets $\Gamma_\nu(K,K) =
2 e\mu^\epsilon K_\nu + \delta\Gamma_\nu^{1L}(K,K)$, with
\beq\label{G1L}
\delta \Gamma^{1L}_\mu(K,K) = -4 \left(e\mu^\epsilon\right)^3 \int[{\nrm d}Q]\,
S_0(Q)S_0(K+Q)\left\{K_\mu + (K_\mu+Q_\mu) K^2 S_0(K+Q)\right\},\eeq
in Feynman gauge.
For $k_0 \equiv i\omega_n \not =0$,
 the one-loop expressions for the vertex and the scalar
self-energy, as given by (\ref{S1L}) and (\ref{G1L}), are
 sufficient to get the order $e^4$  contribution to eq.~(\ref{fullb}):
\beq\label{b41}
\left[\Pi_{\mu\nu}^{(4b)}(0)\right]_{n\ne 0}=-2\left(
e\mu^\epsilon\right)\int[{\nrm
d}K]^\prime\,k_\mu\,\delta\Gamma_\nu^{1L}(K,K)\,S_0^2(K)
\\ \nonumber +8\left(
e\mu^\epsilon\right)^2\int[{\nrm
d}K]^\prime \,k_\mu k_\nu\,\Sigma_{1L}(K)\,S_0^3(K).
\eeq
In the calculation of $\Pi_{00}$ we only have to consider
non vanishing Matsubara frequencies $k_0\ne 0$, so that $\Pi^{(4b)}_{00}$
is completely determined by eq.~(\ref{b41}) above.
The static mode only  contributes to the magnetic sector,
and there the resummation of the thermal masses is again necessary.
In this case, we expand as follows (compare eq.~(\ref{fullpia})):
\beq\label{bexp}
\left[\Pi^{(4b)}_{ii}(0)\right]_{n=0}
= - 2\left(e\mu^\epsilon\right)T\int ({\nrm
d}{\nbf k}) \,k_i \,\delta \Gamma_i({\nbf k}, {\nbf k})
\,{}^*S^2(0,{\nbf k})\\ \nonumber
+8\left(e\mu^\epsilon\right)^2T\int ({\nrm
d}{\nbf k}) \,{\nbf k}^2\,{}^*S^3(0,{\nbf k})\left[\Sigma (0,{\nbf
k})-M^2\right].\eeq
Let $\omega_m$ denote the internal Matsubara frequency in the
one-loop diagrams contributing to $\delta \Gamma_i({\nbf k}, {\nbf k})$ or
to  $\Sigma (0,{\nbf k})$. The non-static modes ($\omega_m\not = 0$)
give rise to mixed two-loop graphs which contribute  to eq.~(\ref{bexp})
only to order $e^5$; they will be ignored in the present
calculation. For the static mode ($\omega_m = 0$), the resummation
of the thermal masses is compulsory, so that the corresponding
self-energy ${}^*\Sigma (0,{\nbf k})$ and vertex-correction
$ {}^*\delta \Gamma_i({\nbf k}, {\nbf k})$ are determined from
dressed one-loop diagrams. The corresponding
result for the self-energy has been  given
in eq.~(\ref{sig*}). For the vertex function, one obtains similarly
\beq\label{gam*}
\left[ k_i {}^*\delta\Gamma_i({\nbf k},{\nbf
k})\right]_{m=0} = -4\left(e\mu^\epsilon\right)^3T\int ({\nrm d}{\nbf q})\,
S_0(0,{\nbf q})\,{}^*S(0,{\nbf k+q})\\ \nonumber\left\{
{\nbf k}^2-{\nbf k}\cdot ({\nbf k+q})({\nbf k}^2-M^2)\,{}^*S(0,{\nbf
k+q})\right\}\\ \nonumber \,=\,  \left(e\mu^\epsilon\right)\,k_i
\,\frac{\del {}^*\Sigma(0, {\nbf k})}{\del k_i}\Biggr|_{m=0}.
\eeq  As shown by the last line above, this result is consistent with the
Ward identity and with the expression (\ref{sig*}) of
  ${}^*\Sigma (0,{\nbf k})$. After combining
 eqs.~(\ref{bexp}), (\ref{sig*}) and (\ref{gam*}),
one obtains the result displayed in  eq.~(\ref{Bii3}) below.

We now summarize the results obtained in this section. We have
\beq\label{a4}
 \Pi^{(4a)}= A_1+A_2+A_3,
\eeq
where
\beq\label{A1}
A_1=2\left(e\mu^\epsilon\right)^2M^2\int[{\nrm
d}K]^\prime \,S_0^2(K)\,,
\eeq
\beq\label{A2}
A_2=-4\left(e\mu^\epsilon\right)^4 \int[{\nrm
d}K]^\prime\,\int[{\nrm d}Q]^\prime\,
S_0(K)S_0(K+Q)S_0(Q)\,,
\eeq
\beq\label{A3}
A_3=2\left(e\mu^\epsilon\right)^4 T^2 \int ({\nrm d}{\nbf k})\int ({\nrm
d}{\nbf q})\, \frac{1}{({\nbf k}^2+M^2)^2}
\\ \nonumber\left\{
\frac{1}{{\nbf q}^2+m^2}+\frac{1}{{\nbf q}^2+M^2}-
\frac{2}{{\nbf q}^2}-\frac{2}{{\nbf q}^2}\frac{{\nbf
k}^2-M^2}{({\nbf q+k})^2+M^2}\right\}.
\eeq
The first two terms, $A_1$ and $A_2$, are obtained from eqs.~(\ref{a41})
and (\ref{S1L}). In writing them, we have omitted
mixed terms with $\omega_n\not = 0$ and $\omega_m= 0$, since they
contribute only to higher orders. The term $A_3$ follows immediately from
eqs.~(\ref{a42}) and (\ref{sig*}).

Similarly
\beq\label{b400}
 \Pi_{00}^{(4b)}(0)= B^1_{00}+B^2_{00}\,,
\eeq
with
\beq\label{B001}
B^1_{00}=8\left(e\mu^\epsilon\right)^2M^2\int[{\nrm d}K]\,k_0^2\,S_0^3(k)\,,
\eeq
\beq\label{B002}
B^2_{00}=-12\left(e\mu^\epsilon\right)^4\int[{\nrm d}K]\int[{\nrm
d}Q]\, k_0^2\, S_0^2(K)\,S_0(K+Q)\,S_0(Q).
\eeq

Finally, for the magnetic contribution we set
\beq\label{b411}
\Pi_{ii}^{(4b)}(0)=B_{ii}^1+B_{ii}^2+B_{ii}^3\,,
\eeq
with
\beq\label{Bii1}
B_{ii}^1=-2\left(e\mu^\epsilon\right)^2M^2\int[{\nrm
d}K]^\prime\,k_i\frac{\del}{\del k_i}S_0^2(K)
\,,\eeq
\beq\label{Bii2}
B_{ii}^2=4\left(e\mu^\epsilon\right)^4\int[{\nrm
d}K]^\prime\,\int[{\nrm
d}Q]^\prime
\,S_0(Q)\,k_i\frac{\del}{\del k_i}\left[ S_0(K) S_0(K+Q)\right]
\,,\eeq
\beq\label{Bii3}
B_{ii}^3=-2\left(e\mu^\epsilon\right)^4T^2 \int ({\nrm d}{\nbf
k})\int({\nrm d}{\nbf q})\, k^i\frac{\del}{\del k^i}
\Biggl\{ \frac{1}{({\nbf k}^2+M^2)^2}\\\nonumber \left[
\frac{1}{{\nbf q}^2+m^2}+\frac{1}{{\nbf q}^2+M^2}-
\frac{2}{{\nbf q}^2}-\frac{2}{{\nbf q}^2}\frac{{\nbf
k}^2-M^2}{({\nbf q+k})^2+M^2}\right]\Biggr\}.
\eeq

One can verify that the magnetic contributions from diagrams (a) and
(b) compensate, as expected.  To see this, note that we can write
\beq
 \Pi_{ii}^{(4a)}=(D-1)\int ({\nrm d}{\nbf k})\,f({\nbf k})
\eeq
and
\beq
\Pi_{ii}^{(4b)}(0)= \int ({\nrm d}{\nbf
k})\,k^i\frac{\del}{\del k^i}f({\nbf k}).
\eeq
The sum of the
two expressions above vanishes after an integration by parts
(allowed by dimensional regularisation).  One can also verify
that $\Pi^{(4a)}_{ii}(0, {\nbf p})+\Pi^{(4b)}_{ii}(0, {\nbf p})$ vanishes as
$p^2$ for $p\to 0$. To do this, it is sufficient to study
the contribution of the static modes for $p_0 =0$ but
 non-vanishing ${\nbf p}$ (that is, to   replace
 $B^3_{ii}$, eq.~(\ref{Bii3}), with the corresponding contribution for
$p\not =0$). Then, it may be verified that, because of the resummation
of thermal masses, the relevant expression admits
a well-defined expansion in powers of $p^2$.

\subsection{Order $e^4$: diagrams 1.c and 1.d}

With bare vertices, and zero external momenta, the contributions coming
from the diagrams
1.c  and 1.d are, respectively,
\beq
\Pi_{\mu\nu}^c(0)=-4\left(e\mu^\epsilon\right)^4\int [{\nrm d}K]
\int [{\nrm d}Q]\, D_{\mu\nu}(Q) S(K) S(K+Q)\,,
\eeq
and
\beq
\Pi_{\mu\nu}^d(0)=-8\left(e\mu^\epsilon\right)^4\int [{\nrm d}K]
\int [{\nrm d}Q]\, D_{\mu\sigma}(Q)\,(2K^\sigma +Q^\sigma
)\,K_\nu \,S^2(K) S(K+Q)\, ,
\eeq
with $k_0=i\omega_n$ and $q_0=i\omega_m$.
(The vertex corrections in these diagrams play no role up to the order $e^4$.)
 If $\omega_n\ne 0$ and
$\omega_m\ne 0$, bare propagators can be used in these
expressions in order to obtain the $e^4$ contribution. If
 $\omega_n=\omega_m=0$, the propagators $^*S$ and
$^*D$ should be used instead.  Finally, the mixed graph where
$\omega_n = 0$ and $\omega_m\ne 0$, or vice-versa, do not contribute
to order $e^4$. In all  cases of interest, the calculation is
straightforward and leads to the following results:
\beq\label{c400}
 \Pi_{00}^{(4c)}(0)=C_{00}^1+C_{00}^2,
\eeq
\beq\label{C001}
C_{00}^1=4\left(e\mu^\epsilon\right)^4\int [{\nrm d}K]\,(1-\delta_{n0})\,
\int [{\nrm d}Q]^\prime\,
S_0(K)\,S_0(K+Q)\,S_0(Q),
\eeq
\beq\label{C002}
C_{00}^2=4\left(e\mu^\epsilon\right)^4\,T^2\int ({\nrm d}{\nbf k})
\int ({\nrm d}{\nbf q})\, \frac{1}{{\nbf q}^2+m^2}\,\frac{1}{{\nbf
k}^2+M^2}\,\frac{1}{({\nbf k+q})^2+M^2}.
\eeq

\beq\label{c4ii}
 \Pi_{ii}^{(4c)}(0)=C_{ii}^1+C_{ii}^2,
\eeq
\beq\label{Cii1}
C_{ii}^1=-4(D-1)\left(e\mu^\epsilon\right)^4\int [{\nrm d}K]^\prime
\int [{\nrm d}Q]^\prime\,
S_0(K)\,S_0(K+Q)\,S_0(Q),
\eeq
\beq\label{Cii2}
C_{ii}^2=-4(D-1)\left(e\mu^\epsilon\right)^4\,T^2\int ({\nrm d}{\nbf k})
\int ({\nrm d}{\nbf q})\, \frac{1}{{\nbf q}^2}\,\frac{1}{{\nbf
k}^2+M^2}\,\frac{1}{({\nbf k+q})^2+M^2}.
\eeq

\beq\label{D00}
\Pi_{00}^{(4d)}(0) =D_{00}=
12\left(e\mu^\epsilon\right)^4\int [{\nrm d}K]
\int [{\nrm d}Q]\,k_0^2\,
S_0^2(K)\,S_0(K+Q)\,S_0(Q).
\eeq

\beq\label{d4ii}
 \Pi_{ii}^{(4d)}(0)=D_{ii}^1+D_{ii}^2,
\eeq
\beq\label{Dii1}
D_{ii}^1=8\left(e\mu^\epsilon\right)^4\int [{\nrm d}K]^\prime
\int [{\nrm d}Q]^\prime\,
[{\nbf k}\cdot ({\nbf
2k+q})]\, S_0^2(K)S_0(K+Q)S_0(Q),
\eeq
\beq\label{Dii2}
D_{ii}^2=8\left(e\mu^\epsilon\right)^4\,T^2\int ({\nrm d}{\nbf k})
\int ({\nrm d}{\nbf q})\, \frac{1}{{\nbf q}^2}\,\frac{{\nbf k}\cdot ({\nbf
2k+q})}{({\nbf k+q})^2+M^2}\,\frac{1}{{\nbf k}^2+M^2}.
\eeq

It can be easily verified that the sum of the magnetic
contributions of diagrams (c) and (d) is a total derivative with
respect to ${\nbf k}$ and therefore vanishes upon  integration. For
instance
\beq
C_{ii}^2+D_{ii}^2=-4\left(e\mu^\epsilon\right)^4\,T^2\int ({\nrm d}{\nbf k})
\int ({\nrm d}{\nbf q})\, \frac{1}{{\nbf q}^2}\,\left[
(D-1)+k^i\frac{\del}{\del k^i}\right]\\ \nonumber
\left( \frac{1}{{\nbf
k}^2+M^2}\,\frac{1}{({\nbf k+q})^2+M^2} \right)\,=\,0,
\eeq
and similarly $C_{ii}^1+D_{ii}^1=0$. One can also verify that the
function $ \Pi_{\mu\nu}^{(4c)}(0,p)+ \Pi_{\mu\nu}^{(4d)}(0,p)$
 is analytic in $p^2$ as
$p\to 0$.

\subsection{Electric mass to order $e^4$}

By adding together the results of sections 3.3 and 3.4, we are now able to
 obtain the correction of order $e^4 T^2$ to $\Pi_{00}(0,0)$. The relevant
equations are (\ref{A1})--(\ref{A3}),  (\ref{B001})--(\ref{B002}),
 (\ref{C001})--(\ref{C002}) and (\ref{D00}), which express the
relevant contributions of the four diagrams in Fig. 1. A simple inspection
of these equations shows that the following compensations arise:
\beq\label{comps} A_2+C^1_{00}=0, \qquad\qquad B_{00}^2+D_{00}=0.\eeq
That is, the only non-trivial contribution of the {\it non-static}
two-loops diagrams  is that given by $A_1+B_{00}^1$, eqs.~(\ref{A1})
and (\ref{B001}). It is proportional to the scalar thermal mass squared
$M^2$:
\beq\label{nonstat}
A_1+B_{00}^1&= &2\left(e\mu^\epsilon\right)^2 M^2 \int[{\nrm
d}K]^\prime \,S_0^2(K)\,\left(1+4 k^2_0\,S_0(K)\right)\\
\nonumber &=&2 M^2 (D-4)\left(e\mu^\epsilon\right)^2\,b_2\, ,\eeq
where the second line follows from the first one after an integration by
parts (see eq.~(\ref{bn})). By evaluating $b_2$ from
  eq.~(\ref{bn}), we  obtain
\beq\label{nonst}
A_1+B_{00}^1\,=\,-\left(\frac{eM}{2\pi}\right)^2\,=\,-\left(
\frac{e^2 T}{4\pi}\right)^2\eeq
for the contribution of the non-static modes.
The contribution to $\Pi^{(4)}_{00}(0,0)$ which arises after  resumming
thermal masses in  purely static two-loops diagrams involves the
 terms $A_3$ and $C^2_{00}$ (eqs.~(\ref{A3}) and (\ref{C002})),
whose sum is computed in Appendix B:
\beq\label{stat}
A_3 + C_{00}^2 &=& -\left(\frac{e^2 T}{4\pi}\right)^2\left(
\frac{m-M}{M} + 4 \ln \frac{m+2M}{2M}\right)\\ \nonumber
&=&  -\left(\frac{e^2 T}{4\pi}\right)^2\left(\frac{2}{\sqrt 3} -1 + 4\ln
\left(1+\frac{1}{\sqrt 3}\right)\right).\eeq
By adding eqs.~(\ref{nonst}) and (\ref{stat}), we derive finally
\beq\label{total4}
\Pi_{00}^{(4)} (0, 0)&=& -\left(\frac{e^2 T}{4\pi}\right)^2\left(
\frac{m}{M} + 4 \ln \frac{m+2M}{2M}\right)\\ \nonumber
&=&  -\left(\frac{e^2 T}{4\pi}\right)^2\left(\frac{2}{\sqrt 3} + 4\ln
\left(1+\frac{1}{\sqrt 3}\right)\right).\eeq
No UV divergencies have been  encountered in
the derivation of this result: all the terms which survive
the compensations (\ref{comps}) are UV finite.
Note that the first line  vanishes if $m=0$, i.e. if we ``forget'' to resum
the static longitudinal photon lines.

 The result (\ref{total4})
represents the main ingredient for computing the correction
of order $e^4$ to the Debye screening mass $m^2_D$.
In order to solve the pole equation (\ref{debyemass}) to this accuracy,
 we actually need the electrostatic polarization operator $\Pi_{00}(0,p)$
for $p^2= - m^2$ up to order $e^4$. That is, we still have to evaluate the
momentum dependence of  $\Pi_{00}(0,p)$ with the prescribed accuracy.
Because of the restriction to soft momenta ($p = i m \sim eT$), this can be
easily done: At one-loop order,
we have $\Pi_{00}^{1L}(0,p)= -m^2 - a \,e^2 p^2 +{\cal O}(e^2 p^4/T^2)$,
with the coefficient $a$ computed in App. A (see
eqs.~(\ref{adef})--(\ref{a})).
At two-loop order,  only the momentum dependence of the {\it static}
diagrams is relevant for  $\Pi_{00}(0,p)$ to order $e^4$
(for non-static graphs, the leading non-trivial behaviour
at small momenta $p\ll T$ is $\sim e^4 p^2$, which is of order $e^6 T^2$
for $p\sim eT$). The only static two loop graph which
is momentum dependent is the diagram 1.c, which
gives the following contribution to   $\Pi_{00}(0,p)$:
\beq\label{C002p}
C_{00}^2 ({p})=4\left(e\mu^\epsilon\right)^4\,T^2\int ({\nrm d}{\nbf k})
\int ({\nrm d}{\nbf q})\, \frac{1}{({\nbf q+p})^2+m^2}\,\frac{1}{{\nbf
k}^2+M^2}\,\frac{1}{({\nbf k+q})^2+M^2}.
\eeq
(For $p=0$, this obviously reduces to $C_{00}^2$, eq.~(\ref{C002}).)
It is easy to see that the momentum dependence of $C_{00}^2 (p)$
{\it does} matter to order $e^4$ if $p\sim eT$:
even if eq.~(\ref{C002p}) is indeed analytic with respect to $p^2$ as
$p\to 0$, the thermal masses which ensure this analitycity
are precisely of order $eT$; it follows that, for momenta $p\sim eT$, all the
terms in the small momentum expansion are of the same order and the exact
expression for $C_{00}^2(p)$ should be used.

When the previous results are set together, we obtain, for $p\sim eT$,
\beq\label{pi4}
\Pi_{00}(0,p)= -m^2 -a\,e^2 p^2 - \delta Z_3 p^2
+\Pi_{00}^{(3)}(0,0)+\Pi_{00}^{(4)}(0,0)
\\ \nonumber +\left(C_{00}^2 ({p}) - C_{00}^2 (0)\right)
+{\cal O}(e^5T^2).
\eeq
The counterterm
\beq \delta Z_3 = -\frac{e^2}{3(4\pi)^2}\,\frac{1}{\epsilon} \eeq
 accounts for the photon wave-function renormalization
in the minimal substraction scheme and cancels
the divergent piece
of $a$ in the limit $D\to 4$ (see eq.~(\ref{a})).
This leaves the following renormalised value for the coefficient $a$:
\beq\label{aR}
a_R=\frac{1}{3(4\pi)^2}\left(\gamma+2+\ln\frac{\mu^2}{4\pi
T^2}\right)
\eeq
To order $e^4$, the solution to eq.~(\ref{debyemass}) is
\beq\label{sol}
m_D^2=m^2(1-a_Re^2)-\Pi_{00}^{(3)}(0,0)-\Pi_{00}^{(4)}(0,0)
-\left(C_{00}^2 ( im) - C_{00}^2 (0)\right)+{\cal O}(e^5T^2).
\eeq
 The difference  $C_{00}^2(p=im)- C_{00}^2 (0)$
is
readily computed from eqs.(\ref{C002p}) and (\ref{C002}) as
\beq\label{diff}
C_{00}^2 ( p=im) - C_{00}^2 (0) = \left(\frac{e^2 T}{2\pi}\right)^2\,
\int_0^\infty
\frac{{\nrm d}x}{x}\,{\nrm e}^{-(2M+m)x}\left(\frac{{\nrm sinh} \,mx}{mx}-1
\right) \\ \nonumber
 =  \left(\frac{e^2 T}{2\pi}\right)^2\,\left(1\,-\,\frac{M}{m}\,\ln\frac{2M+m}
{2M}\,-\,\frac{M+m}{m}\,\ln\frac{2(M+m)}{2M+m}\right).\eeq
By also using eqs.~(\ref{total3}) and (\ref{total4}) for
 $\Pi_{00}^{(3)}(0,0)$ and $\Pi_{00}^{(4)}(0,0)$, we finally obtain
\beq\label{mD4}
m_D^2\,= \,T^2\left(\frac{e^2}{3}-\frac{e^3}{4\pi}
+\,b\,\frac{e^4}{(2\pi)^2}
-\frac{e^4}{(12\pi)^2}\left[\gamma+2+\ln \frac{\mu^2}{4\pi
T^2}\right]\right)\,+\,{\cal O}(e^5T^2),
\eeq
where $b$ is the positive number
\beq
b\,=\,-1+ \frac{1}{2\sqrt 3}+\left(1+\frac{\sqrt 3}{2}
\right)\ln \left(1+\frac{2}{\sqrt 3}\right)\,=\,0.7211328...\, ,\eeq
and $e$ is the gauge coupling at the scale $\mu$ ($e\equiv e(\mu)$),
as defined by the minimal substraction scheme.
The expression (\ref{mD4}) is independent of the choice of $\mu$ if
${\nrm d}e/{\nrm d}\ln \mu=e^3/48\pi^2$, which is precisely
 the one-loop  $\beta-$function  of scalar QED.
To verify that the screening mass (\ref{mD4})
 is actually gauge independent, we rely on the
gauge invariance of $\Pi_{00}(0,0)$
and on the fact that the momentum dependent terms which enter  $\Pi_{00}(0,p)$
to order $e^4$ (see eq.~(\ref{pi4})) arise from gauge-independent diagrams.
(The latter assertion is obvious for the one-loop contribution, $\Pi^{1L}_{00}
(0,p)$; as for the two-loop contribution of eq.~(\ref{diff}), this
only involves the time-time piece of the static photon propagator,
which is the same in all covariant gauges.)

\subsection{Summary: effective theory for static modes}

In the previous sections, we have computed the zero-momentum limit
($\omega=0$, $p\to 0$) of the photon polarization tensor to order $e^4$.
There are two type of terms
occuring in the perturbative expansion, and we want to discuss this
in more detail now. Consider the electric piece,
 $\Pi_{00}(0,0)$, as  given to order $e^4$ by eqs.~(\ref{HTL}),
(\ref{total3}) and (\ref{total4}). Some of the contributions
in these equations arise from
{\it purely non-static} one- and two-loop graphs, which are computed with the
standard Feynman rules (i.e. without resummation), except for the fact
that the terms with $\omega_n = 0$ are excluded from
 the sums over  Matsubara frequencies.
 This   is the case for the
leading order electric mass, (\ref{HTL}), and also for the two-loop
 contribution of eq.~(\ref{nonst}).
Since no IR problem arises in such diagrams, naive power counting applies.
Quite generally then,  $n$-loop  diagrams with no static internal line
will contribute to $\Pi_{00}(0,0)$ to order $e^{2n}$.
The second type of terms involve one- and two-loop  diagrams in which
{\it all} the internal lines are static and include the corresponding
 thermal masses.
The non-analytic contribution of order $e^3$, eq.~(\ref{total3}), belongs to
 this category, but this is also the case for the order-$e^4$ contribution
of eq.~(\ref{stat}). As already mentioned, the {\it mixed} two-loop graphs
(where one of the internal frequencies is vanishing, and the other is not)
contribute to order $e^5$. In general, it can be
verified by power counting that a $n$-loop diagram ($n\ge 1$) with only
static (dressed) internal lines
contributes to $\Pi_{00}(0,0)$ to the order $e^{n+2}$, and  eventually
dominates over the corresponding non-static contribution (of order
$e^{2n}$) as soon as $n\ge 3$. As for the mixed $n$-loop graphs
(with $n\ge 2$), their leading contribution to $\Pi_{00}(0,0)$ is also of
the order $e^{n+2}$ for $n\ge 3$ (it is only of the order $e^5\equiv
e^{n+3}$ for $n=2$).

A different, probably more transparent, way to look at
the perturbative expansion  after the  resummation of the
thermal masses is to consider, as an intermediate step,
the effective three-dimensional theory for static and long wavelength
($p\simle eT$) fields. This is obtained
after integrating  non-static loops with static external
lines\cite{Nadkarni83}.
The lagrangian of the effective  theory reads ($D_i\equiv \del_i - i e_3 A_i$)
\beq\label{L3}
{\cal L}_{eff}=\frac{1}{4}F_{ij}^2 + \frac{1}{2\alpha}(\del_i A_i)^2
+\frac{1}{2} \, (\del_i A_0)^2+\frac{1}{2} \,m_0^2\,A_0^2
+e_3^2 A_0^2 \phi^\dagger\phi\\ \nonumber
+(D_i\phi)^\dagger(D_i\phi) + M_0^2 \phi^\dagger\phi +\delta {\cal L}.\eeq
The magnetostatic gauge fields $A_i({\nbf x})$, $i=1,\,\,2,\,\,3$,
the electrostatic (gauge-invariant) field $A_0({\nbf x})$, and the
scalar field $\phi({\nbf x})$,
may be identified, up to normalizations,
with the zero-frequency components of the original fields.
The term $\delta {\cal L}$ contains higher order vertices, but also
 derivative corrections to the $n$-point
vertices shown explicitly; in particular,
it contains the counterterms necessary for UV renormalization. All these
operators,  as well as the parameters $e_3$, $m_0^2$ and $M_0^2$,
are obtained as power series in $e^2$ by evaluating non-static diagrams with
static external lines in the original theory, and by expanding with respect
to the external momenta. To leading order, $e_3^2\approx e^2 T$, $m_0^2
\approx m^2$ and $M_0^2\approx M^2$. Generally, a vertex with $n$ static
external lines which   is absent at the tree-level
 is first induced at the order $e^n$, via non-static one-loop
graphs. Actually, some particular vertices may be induced at a level
higher than $e^n$, or may even vanish, because of some symmetry.
 The lagrangian (\ref{L3}) is invariant under static
gauge transformations.

The loop corrections in the effective theory generate
all the static diagrams of the original theory.
They also include the {\it mixed} graphs: indeed, a non-static subgraph
of an original mixed diagram appears as a bare vertex in the effective
 theory, while a static subgraph appears as  a loop correction.
The calculation that we have done before can be understood simply
in terms of this reorganized perturbation theory. Consider for exemple
 the calculation of $\Pi_{00}(0,0)$. In the effective theory, we have
$\Pi_{00}(0,0)\equiv - m^2_0-\Sigma_{A_0}({\nbf p}=0)$,
v where $\Sigma_{A_0}({\nbf p})$ is the
self-energy of the field $A_0$ in the effective theory.
Thus, to get $\Pi_{00}(0,0)$ to order $e^4$, we need both $m_0^2$ and
 $\Sigma_{A_0}(0)$ to order $e^4$. We already know the result for
$m_0^2$: by evaluating the corresponding non-static
one- and two-loop graphs, we have obtained
(see eqs.~(\ref{HTL}) and (\ref{nonst}))
\beq\label{m0}
m_0^2=\frac{e^2T^2}{3}\left\{1+\frac{3}{(4\pi)^2}\,e^2\,+{\cal O}(e^4)
\right\}\,.\eeq
To order $e^4$,  $\Sigma_{A_0}(0)$ is given
by the one- and two-loop diagrams in Fig. 8, where the leading order
 vertices ($e_3^2 = e^2 T$) and massive propagators  should be used ($M_0^2
= M^2$ and  $m_0^2=m^2$). Their evaluation is straightforward, and leads  to
\beq\label{Sig0}
\Sigma_{A_0}(0)=-\frac{e^3 T^2}{4\pi}
+\left(\frac{e^2 T}{4\pi}\right)^2\left(\frac{2}{\sqrt 3} -1 + 4\ln
\left(1+\frac{1}{\sqrt 3}\right)\right),\eeq
 which reproduces the contribution of the relevant
 static graphs in the original theory.
By adding the contributions (\ref{m0}) and (\ref{Sig0}),
 we recover our previous result for $\Pi_{00}(0,0)$.

It is  easy to see the systematics of the
  higher order corrections.  The perturbative expansion of
$\Pi_{00}(0,0)$ decomposes into the sum of two terms: the first one,  $m_0^2$,
is a power series in $e^2$ and
arises from the non-static diagrams of the original theory; the second term,
$\Sigma_{A_0}({\nbf p}=0)$, is obtained as
 a power series in $e$ in the effective
theory. The fact that it is $e$ rather than $e^2$ which governs the
perturbative expansion in the effective theory is due to the fact that
the loop integrals in the effective theory would be IR divergent
in the absence of the thermal masses.  Thus,
 a $n$-loop diagram contributing to  $\Sigma_{A_0}$ has an
explicit factor of $e_3^{2n}\sim e^{2n} T^n$; since it has dimension two,
it is proportional to $e^{2n} T^2 (T/M)^{n-2}
\sim (e^4 T^2) e^{n-2}$, where $M\sim eT$ is any of the
 bare masses $m_0$ or $M_0$, or  the external momentum, playing the role
of an infrared cut-off. Since
$n\ge 1$, we see emerging a power expansion in $e$, starting at the
order $e^3$. We have assumed here that all the vertices entering the
$n$-loop diagram are of the type explicitly depicted in eq.~(\ref{L3}),
i.e. vertices which are already present in the original theory at
 the tree-level. Actually, the
 new vertices contained in ${\delta {\cal L}}$ will
also contribute, but only starting at the order $e^5$ (as may be verified
by power counting). In particular, to obtain the electric screening mass
to order $e^5$ we need the bare parameters 
 $m_0^2$ and $M_0^2$ to two-loop order in the original theory
(e.g. eq.~(\ref{m0}) for $m_0^2$)
 and the corrections to  $\Sigma_{A_0}({\nbf p}=0)$
up to three-loop order in the effective theory. The latter will involve
one-loop diagrams with some of the 4-point effective vertices
from  ${\delta {\cal L}}$.

\setcounter{equation}{0}
\section{Conclusion}

In this paper we computed  directly the Debye screening masses for QED
and SQED two orders above their lowest nontrivial values
 and verified explicitly that the
corresponding magnetic  masses vanished in the same approximation.
It was also shown that the  results were both gauge and
renormalization group invariant. For QED, we further argued that
the vanishing of the magnetic mass holds to all orders in  perturbation
theory: Since the  (odd)  Matsubara frequencies of the fermionic propagators
play the role of a mass, the static photon lines are always coupled
to massive fields; there is no closed loop of massless fields. This is
 enough to ensure the analyticity with respect to $p^2$ ($p$ being the
external three momentum)
 of the  Feynman diagrams contributing to the static polarisation tensor.
These arguments can be extended
to SQED as well, in spite of the fact that the propagator of the scalar
particle may be at zero Matsubara frequency:
Once thermal masses of order $gT$  are given
to the charged particles  and to the longitudinal  photons by the
appropriate resummations, the situation in SQED becomes similar to
that of QED ---- the magnetostatic fields, which remain massless,
couple only to massive particles. By contrast, the
properties of the polarisation tensor are different in non-abelian
theories. For example, in QCD there is  no identity like
(\ref{mag}) and, in fact, the self-interactions of the  magnetostatic
 gluons are believed to generate a magnetic screening mass $\sim g^2T$
\cite{GPY}.

In QED, the above arguments on the
 analyticity, with respect to $p^2$, of the static polarisation tensor
are  valid for momenta
 $p$ smaller than $T$ because the odd fermionic frequencies $\sim T $
provide the relevant infrared cut-off. This is why, for example, in
eq. (3.21) we could use the small momentum  expansion of $\Pi_{00}(0,p)$
up to $p\sim im$. On the other hand, in SQED, the infrared cut-off is
provided by the screening masses
$\sim gT$, and the small $p^2$ expansion holds only for $p<gT$. Thus for
example, in the calculation of the electric mass at order $e^4$, we could not
expand $\Pi_{00}(0,p)$ for $p\sim im$ (see eq.(4.65)). In fact
singularities do occur for imaginary values of
$p$ of order
$gT$. For exemple,
${}^*\Pi_{ii}(0, p)$ in eq.~(\ref{iiop}) has  branch point
singularities at $p = \pm 2iM$, and
${}^*\Sigma(0, p)$, eq.~(\ref{SIG*}), has  similar singularities
for  $p = \pm iM$. It is interesting to note, however,  that while
${}^*\Pi_{ii}(0, p= \pm 2iM)=0$,
${}^*\Sigma(0, p= \pm iM)$ diverges. This latter behaviour
prevents the calculation of the correction of order $e^3 T^2$ to the
screening mass $M^2$ of the scalar field
along the lines followed in this paper for $m_D^2$.

An  analytical structure similar to that of the scalar self-energy
${}^*\Sigma(0, p)$ has been observed in the  calculation of the
non-abelian Debye mass at the next to leading order, i.e. $m_D^2$ up to order
$g^3 T^2$
  \cite{Rebhan93,BN94}. This analogy between the scalar self-energy
in SQED and the gluon self-energy in QCD can be expected from
 the similarity between the charged scalar sector of ${\cal L}_{eff}$,
eq.~(\ref{L3}), and the electrostatic sector of the corresponding
three-dimensional effective theory for high temperature
 QCD\cite{Nadkarni83,Braaten94}. In Refs.   \cite{Rebhan93,BN94}, the
logarithmic divergence in the electrostatic
 gluon polarisation tensor at $p = im$
 has been cut-off by giving
the static gluon a magnetic mass of order
$g^2 T$. Because of the analogous difficulty in SQED where no magnetic
mass can be generated, we feel that it is worthwhile to look for
alternative treatement of this problem\cite{wip}.

We had stated that the derivation of (\ref{derpress}) was formal
but checked its correctness  explicitly to fifth order in QED.
On the other hand if one has faith in that identity, then
our verification may be reinterpreted as
actually having provided a nontrivial check on the
correctness of our resummed perturbation expansion.
For SQED we did not verify the
identity to the fourth order but in this case we repeated our computations
using two different resummation schemes, thus providing again
some useful cross checks.

Of course, static correlators are not the only relevant probes of a plasma.
The plasma frequency and damping rate are two important
quantities which can be deduced only from dynamic correlation functions.
Compared to static quantities, the computation of dynamic quantities
is more involved --- not only must one work in real-time (that is an
analytical continuation of imaginary time, or the real-time formalism
itself), but one also has to use the full machinery developed in \cite{BP90}
in the resummation necessary for higher order calculations.
 As far as we know,
of the four possible explicit resummation methods mentioned in the
Introduction, only method (c) (use of rearranged lagrangians incoporating
the hard thermal loop effects) has so far been used for dynamical
calculations beyond leading order \cite{BP90,RP92,Schulz,Kraemmer94}.\\

\vspace*{0.3cm}
\noindent {\bf Acknowledgement\\}
R.P. thanks the Theoretical Physics group at the TIFR, Bombay, for
hospitality during a visit at the final stages of this work.\\

\setcounter{equation}{0}
\vspace*{2cm}
\renewcommand{\theequation}{A.\arabic{equation}}
\appendix{\noindent {\large{\bf Appendix A}}}

In this Appendix, we review a derivation
of the hard thermal loop for the photon polarization tensor
which emphasizes its classical character and the ensuing universality
of its structure for any gauge theory.
To start with, we recall that $\Pi_{\mu\nu} = \delta j^{ind}_\mu/
\delta A^\nu$, where $j_\mu^{ind}$ is the current induced by the
electromagnetic field. To leading order in $e$
 \beq\label{jind}
j^{ind}_\mu(x)=e\int\frac {{\nrm d}^4K}{(2\pi)^4}\,2k^\mu\,W(K,x)=
e\int\frac {{\nrm d}^3k}{(2\pi)^3}\,\frac{k^\mu}{\epsilon_k}\left[
\delta N_+({\nbf k}, x)-\delta N_-({\nbf k}, x)\right],\eeq
where
$W(K,x)=2\pi\,\delta(k_0^2-\epsilon_k^2)\,\left[ \theta(k_0)
\delta N_+({\nbf k}, x) + \theta(-k_0)
\delta N_-({-\nbf k}, x)\right]$ is a gauge-invariant Wigner function
 for the charged particles,  ultimately related to the scalar
propagator in the presence of the gauge fields\cite{BI}. The $\delta$-function
defines the mass-shell for the scalar particles; in the present
approximation,  it coincides
with the {\it free} mass-shell, $k_0^2=\epsilon_k^2$ (with $\epsilon_k = k$
for massless particles). The equation satisfied by the Wigner functions
$\delta N_\pm({\nbf k}, x)$ is\cite{BI}
\beq\label{eqN}
\left( v\cdot \del_x\right )\,\delta N_\pm({{\nbf k}},x)=\mp\, e\,
{\nbf v\cdot E}(x)\,\frac {dN_0}{d \epsilon_k},\eeq
where $v^\mu \equiv (1, {\nbf v})$, ${\nbf v}\equiv
 {\nbf k}/\epsilon_k$,
  $v\cdot\del_x=\del_t+{\nbf v}\cdot {\bfgrad}$, ${\nbf E}$ is the
mean electric field, and
$N_0({\epsilon_k})=  1/(\exp(\beta\epsilon_k)-1)$  is the equilibrium
 occupation factor for bosons. Note that for an abelian plasma this
is merely the (linearized) Vlasov
 equation. By combining eqs.~(\ref{jind})
 and (\ref{eqN}), one finds $j^{ind}_\mu(\omega, {\nbf p})=
\Pi_{\mu\nu}(\omega, {\nbf p})\,A^\nu(\omega, {\nbf p})$, with
\beq\label{kinpi}
\Pi_{\mu\nu}(\omega, {\nbf p})\,=\,
2e^2 \int\frac {{\nrm d}^3k}{(2\pi)^3}\,\frac{d N_0}{d\epsilon_k}\left(
\delta_{\mu 0}\delta_{\nu 0} - v_\mu v_\nu\,\frac{\omega}{\omega - {\nbf
v \cdot p}}\right ).\eeq
For $\epsilon_k=k$, the radial integral over $k$ can be easily performed,
and one recovers the expression (\ref{Pihtl}).

In abelian
plasmas, $j^{ind}_\mu$ is gauge-invariant and linear in $A_\mu$,
as shown by eqs.~(\ref{jind}) and (\ref{eqN}),  so that  there is no
 hard thermal loop for the $n$-photon vertices with $n > 2$.
In contrast,  the QCD induced current is a
 {\it color} vector, i.e. $j^a_\mu$ transforms
 in the adjoint representation of $SU(N)$. Then, as required by
 gauge symmetry, the kinetic equation (\ref{eqN})
involves a covariant line-derivative
(i.e. $v\cdot\del_x$ is replaced by $v\cdot D_x$ in its l.h.s.),
so that the color current is a {\it non-linear} functional
of the gauge potentials. It follows then that the higher functional derivatives
of $j^a_\mu$ with respect to $A_\mu^a$ are also non-vanishing, and define
hard thermal loops for vertex functions with an arbitrary number
of soft external gluon lines. However, for weak fields,
$|A_\mu|\to 0$, the kinetic equation (\ref{eqN}) is formally the same
for abelian and non-abelian plasmas. This explains why the expressions
 obtained for the polarization tensor in this approximation
are, up to a trivial factor  which counts the relevant
degrees of freedom,   identical for QED, QCD or scalar QED.

As for the hard thermal loop for the scalar self-energy,
this describes the response of the plasma to long-wavelength scalar
mean fields. Since according to eq.~(\ref{Sihtl}), $\Sigma^{(2)}$ is
 a {\it local} operator, gauge covariance requires this operator to be
 independent of the gauge mean field\cite{BI},
so that there are no  hard thermal loops for vertices
involving scalar and photon external lines. We conclude that
the self-energy corrections in eqs.~(\ref{Pihtl}) and (\ref{Sihtl})
are the {\it  only} hard thermal loops  for  SQED, in accordance
with the power counting arguments of Refs. \cite{BP90,Kraemmer94}.

The previous derivation of $\Pi_{\mu\nu}$
 can be generalized to take into account the thermal mass
 $M$ aquired by the charged particles. Since $M\ll T$, the only  effect is the
modification of the mass shell condition which becomes $k^2_0=
\epsilon_k^2\equiv k^2 +M^2$. After inserting this into  eq.~(\ref{kinpi}),
one obtains
\beq\label{kin3}
\Pi_{\mu\nu}(\omega=0, {\nbf p})\,=\,-m^2(T,M)
\,\delta_{\mu 0}\delta_{\nu 0},\eeq
with
\beq
m^2(T,M) =  -\, 2e^2 \int\frac {{\nrm d}^3k}{(2\pi)^3}\,
\frac{d N_0}{d\epsilon_k}\,=\,\frac{e^2T^2}{3} \,-\,\frac{e^2 MT}{4\pi}\,
+{\cal O} \left({e^2 M^2}\right).\eeq
As advertised in section 3.2, this
is the correct electric mass up to the order $e^3$.

\setcounter{equation}{0}
\vspace*{2cm}
\renewcommand{\theequation}{B.\arabic{equation}}
\appendix{\noindent {\large{\bf Appendix B}}}

In this appendix, we derive some formulae which are referred to in Sect. 3.
We  consider first the small momentum behaviour of the one-loop static
polarization tensor, $\Pi^{1L}_{\mu\nu}(0, {\nbf p})$. For the magnetic sector,
the diagrams in Fig. 3 imply
\beq\label{1Lii}
\Pi_{ii}^{1L}(0, {\nbf p})=
 \left(e\mu^\epsilon\right)^2 \int[{\nrm d}K]
 \,\left\{\frac{2(D-1)}{\omega_n^2+{\nbf k}^2}-
\frac{({\nbf 2k+p})^2}{(\omega_n^2+{\nbf k}^2)
\left(\omega_n^2+({\nbf k+p})^2\right)}\right\}.\eeq
The dominant contribution as $p\to 0$ is given
by the $\omega_n=0$ term in the Matsubara sum and it is linear in $p$:
\beq\label{1Liiexp}
\left[\Pi_{ii}^{1L}(0, {\nbf p})\right]_{n=0}&=&
 \left(e\mu^\epsilon\right)^2\,{\nbf p}^2\,T\int({\nrm d}{\nbf k}) \,
\frac{1}{{\nbf k}^2}\,\frac{1}{({\nbf k+p})^2}\\ \nonumber
&=& \frac{e^2 p T}{4\pi^2}\int_0^\infty \frac{dx}{x}\,\ln
 \biggl|\frac{x+1}{x-1} \biggr|\,=\,\frac{1}{8}\,e^2 pT .\eeq
For the time-time component we obtain
\beq\label{1L00}
\Pi_{00}^{1L}(0, {\nbf p})&=&
- \left(e\mu^\epsilon\right)^2\int[{\nrm d} K]
\,\left\{\frac{2}{\omega_n^2+{\nbf k}^2}-
\frac{4\omega_n^2}{(\omega_n^2+{\nbf k}^2)
\left(\omega_n^2+({\nbf k+p})^2\right)}\right\}\\ \nonumber
&=& - m^2 + \left(e\mu^\epsilon\right)^2 \int[{\nrm d} K]
\frac{4\omega_n^2}{(\omega_n^2+{\nbf k}^2)}
\left(\frac{1}{\omega_n^2+({\nbf k+p})^2}-
\frac{1}{\omega_n^2+{\nbf k}^2}\right),\eeq
where
\beq\label{m}
m^2\equiv -\Pi^{1L}_{00}(0,0)= 2(D-2)\left(e\mu^\epsilon\right)^2
\int[{\nrm d}Q] \,S_0(Q)= \frac{e^2 T^2}{3}.\eeq
Note that the static mode $\omega_n=0$ does not contribute to the
last sum in eq.~(\ref{1L00}); accordingly, one can expand the denominator
for small  $p$  without generating IR  singularities:
\beq\label{denexp}
\frac{1}{\omega_n^2+({\nbf k+p})^2}=\frac{1}{\omega_n^2+{\nbf k}^2}\left\{
1-\frac {{\nbf p}^2 + 2{\nbf k\cdot p}}{\omega_n^2+{\nbf k}^2} +
\frac {\left({\nbf p}^2 + 2{\nbf
 k\cdot p}\right)^2}{\left(\omega_n^2+{\nbf k}^2\right)^2}
\, - \, ... \right\}.\eeq
By  keeping only the terms quadratic in $p$, we obtain
\beq\label{1L00exp}
\Pi_{00}^{1L}(0, {\nbf p})&=&-m^2 - a \,e^2\,p^2 +{\cal O}(e^2 p^4/T^2),\eeq
where
\beq\label{adef}
a&\equiv &  4 \mu^{2\epsilon} \int[{\nrm d} K]
\frac{\omega_n^2}{(\omega_n^2+{\nbf k}^2)^3}
\left(1- \frac{4}{D-1}\,\frac{{\nbf k}^2}{\omega_n^2+
{\nbf k}^2}\right)\\ \nonumber
&=& \frac {5-D}{3} \mu^{2\epsilon} \int[{\nrm d} K]^\prime
S_0^2(K).\eeq
and some integrations by parts have been necessary to get
the last line. By also using eq.~(\ref{bn}) with $n=2$
and by expanding in $\epsilon$, we obtain
\beq\label{a}
a=\frac{1}{3(4\pi)^2}\left(\frac{1}{\epsilon}\,+\gamma+2+\ln\frac{\mu^2}{4\pi
T^2} +{\cal O}(\epsilon)\right).\eeq

Finally, we turn to the evaluation
of the integrals (\ref{A3}) and (\ref{C002}) which contribute to
$\Pi_{00}(0)$ to order $e^4$. We organize this computation as follows:
\beq\label{AC}
A_3+C_{00}^2= a_1 + a_2 + a_2,\eeq
where
\beq\label{a1}
a_1\equiv 2\left(e\mu^\epsilon\right)^4 T^2 \int ({\nrm d}{\nbf k})\int ({\nrm
d}{\nbf q})\, \frac{1}{({\nbf k}^2+M^2)^2}\left(
\frac{1}{{\nbf q}^2+m^2}+\frac{1}{{\nbf q}^2+M^2}-
\frac{2}{{\nbf q}^2}\right),\eeq
\beq\label{a2}
a_2\equiv 8\left(e\mu^\epsilon\right)^4 M^2
T^2 \int ({\nrm d}{\nbf k})\int ({\nrm
d}{\nbf q})\, \frac{1}{{\nbf q}^2}\,
\frac{1}{({\nbf k}^2+M^2)^2}\,\frac{1}{({\nbf q+k})^2+M^2},\eeq
and
\beq\label{a3}
a_3\equiv -4\left(e\mu^\epsilon\right)^4 T^2 \int ({\nrm d}{\nbf k})\int ({\nrm
d}{\nbf q})\, \frac{1}{{\nbf k}^2+M^2}\,\frac{1}{({\nbf q+k})^2+M^2}\,
\left(\frac{1}{{\nbf q}^2}-\frac{1}{{\nbf q}^2+m^2}\right).\eeq
All the integrals above are UV and IR finite
in $D=4$, and thus we can set $\epsilon =0$.
The evaluation of $a_1$ is straightforward, with the result
\beq\label{a11}
a_1=  -\left(\frac{e^2 T}{4\pi}\right)^2\,\frac{m+M}{M}.\eeq
In order to compute $a_2$ and $a_3$, it is convenient to use the
coordinate-space representation of the static propagators,
by writing
\beq\label{coul}
\frac{1}{k^2+M^2} = \int {\nrm d}^3x\,{\nrm e}^{i{\vec k\cdot \vec x}}\,\,\frac
{{\nrm e}^{-Mx}}
{4\pi x},\qquad\qquad
\frac{1}{(k^2+M^2)^2} = \frac {1}{8\pi M}
\int {\nrm d}^3x\,{\nrm e}^{i{\vec k\cdot \vec x}}\,{{\nrm e}^{-Mx}},\eeq
with $x\equiv |{\vec x}|$. Then one obtains
\beq\label{a21}
a_2 = 2 \left(\frac{e^2 T}{4\pi}\right)^2,\eeq
and
\beq\label{a31}
a_3 = -4 \left(\frac{e^2 T}{4\pi}\right)^2\,\int_0^\infty
\frac{{\nrm d}x}{x}\left({\nrm e}^{-2Mx}- {\nrm e}^{-(2M+m)x}\right)
=-4 \left(\frac{e^2 T}{4\pi}\right)^2\,\ln\,\frac{m+2M}{2M}.\eeq
By adding together eqs.~(\ref{a11}), (\ref{a21}) and (\ref{a31}),
 we obtain the expression (\ref{stat}).

  \newpage

\begin{center}
{\Large\bf Figure captions}
\end{center}
\vspace*{1cm}
\noindent Figure 1. The skeleton diagrams for the self-energy
 of the photon in scalar QED.

\vspace*{1cm}
\noindent Figure 2. The two loop  contributions to the
 photon self energy in QED.

\vspace*{1cm}
\noindent Figure 3. Two loop diagrams in QED in which the photon internal
line is dressed with the thermal mass.

\vspace*{1cm}
\noindent Figure 4. One loop contributions to the polarisation tensor in SQED.

\vspace*{1cm}
\noindent Figure 5. One loop contributions to the scalar self-energy in
SQED.

\vspace*{1cm}
\noindent Figure 6. One loop contributions to the polarisation tensor in
SQED, with the static
internal lines dressed by thermal masses.

\vspace*{1cm}
\noindent Figure 7. One loop contributions to the scalar self-energy in
SQED, with the static
internal lines dressed by thermal masses.

\vspace*{1cm}
\noindent Figure 8. The self-energy of the static photon in the
three-dimensional
 efffective theory (see eq.(4.70)). Diagram (c) gives no contribution
 in dimensional regularisation. Dashed line: longitudinal photon.
Wavy line: transverse photon. Full line: scalar field.

\end{document}